\pdfoutput=1
\documentclass[12pt]{article}
\usepackage[
top = 2.5cm, 
bottom = 2.5cm,  
left = 2.55cm,  
right = 2.55cm]{geometry}
\usepackage{enumitem}
\usepackage{subcaption}
\usepackage{caption}
\usepackage{latexsym} 
\usepackage{verbatim}
\usepackage{tikz}
\usetikzlibrary{matrix}
\usepackage{color}
\usepackage{graphicx,amssymb,amsfonts,amsmath,amssymb,amscd,amstext, mathrsfs}
\usepackage{graphicx} 
\usepackage{bbm}
\usepackage{dsfont}
\usepackage{xcolor}
\usepackage[colorlinks=true, citecolor=black, linkcolor=black, allcolors=black]{hyperref}   
\usepackage{amsmath}
\usepackage{empheq}
\usepackage{hyperref}
\usepackage{cite}

\newlength\dlf

\def\l{\lambda}
\def\o{\over}
\def\D{\Delta}
\def\gap{{\rm gap}}
\def\rar{\rightarrow}
\def\i{\infty}
\def\foot{\footnote}

\newcommand{\bw}{\begin{widetext}}
\newcommand{\ew}{\end{widetext}}
\newcommand{\bea}{\begin{eqnarray}}
\newcommand{\eea}{\end{eqnarray}}
\newcommand{\be}{\begin{equation}}
\newcommand{\ee}{\end{equation}}
\newcommand{\nn}{\nonumber}

\renewcommand{\bar}[1]{\overline{#1}}

\newcommand{\<}{\langle}
\renewcommand{\>}{\rangle}

\renewcommand{\cal}{\mathcal}
\newcommand{\CO}{\mathcal{O}}

\newcommand{\CL}{\mathcal{L}}

\renewcommand{\O}{\mathcal{O}}

\DeclareMathOperator*{\res}{Res}

\DeclareFontShape{OT1}{cmr}{mx}{n}{<->cmr10}{}
\newcommand{\titlefont}{\fontseries{mx}\selectfont}

\def\frac#1#2{{#1\over #2}}

\begin{document}

\begin{titlepage}

\begin{flushright} 
\end{flushright}

\begin{center} 

\vspace{0.35cm}

{\fontsize{21.5pt}{0pt}{\titlefont
Model-Dependence of  Minimal-Twist OPEs  \\  
 \vspace{0.4cm}  
 in $d>2$  Holographic CFTs
}}

\vspace{1.6cm}  

{{A. Liam Fitzpatrick$^1$, Kuo-Wei Huang$^1$, David Meltzer$^{2}$, \\
\vspace{0.2cm}
Eric Perlmutter$^{2}$,  David Simmons-Duffin$^2$}}

\vspace{1cm} 

{{\it
$^1$Department of Physics, Boston University, 
Boston, MA  02215
\\
\vspace{0.1cm}
$^2$Walter Burke Institute for Theoretical Physics, 
Caltech, Pasadena, CA 91125
}}\\
\end{center}
\vspace{1.5cm}

{\noindent Following recent work on heavy-light correlators in higher-dimensional conformal field theories (CFTs) with a large central charge $C_T$, we clarify the properties of stress tensor composite primary operators of minimal twist, $[T^m]$, using arguments in both CFT and gravity. We provide an efficient proof that the three-point coupling $\langle \mathcal{O}_L\mathcal{O}_L [T^m]\rangle$, where $\mathcal{O}_L$ is any light primary operator, is independent of the purely gravitational action.  Next, we consider corrections to this coupling due to additional interactions in AdS effective field theory and the corresponding dual CFT. When the CFT contains a non-zero three-point coupling $\langle  TT \mathcal{O}_L\rangle$, the three-point coupling $\langle \mathcal{O}_L\mathcal{O}_L [T^2]\rangle$ is modified at large $C_T$ if $\langle TT\mathcal{O}_L \rangle \sim \sqrt{C_T}$. This scaling is obeyed by the dilaton, by Kaluza-Klein modes of prototypical supergravity compactifications, and by scalars in stress tensor multiplets of supersymmetric CFTs. Quartic derivative interactions involving the graviton and the light probe field dual to $\mathcal{O}_L$ can also  modify the minimal-twist couplings; these local interactions may be generated by integrating out a spin-$\ell \geq 2$ bulk field at tree level, or any spin $\ell$ at loop level. These results show how the minimal-twist OPE coefficients can depend on the higher-spin gap scale, even perturbatively.}

\end{titlepage}

\tableofcontents


\section{Introduction 
}
 
Much of physics is concerned with finding the simplest description possible that nevertheless captures some universal behavior of interest. Ideally, such descriptions can be systematically improved. Effective field theories, and the emergent universality at low energies that they describe, provide a robust instance of this approach.  Holographic descriptions of strongly coupled physics, through the AdS/CFT correspondence \cite{Maldacena:1997re, Witten:1998qj, Gubser:1998bc}, may be cleanly derived as applications of effective field theory in the bulk, provided there exists a certain separation of scales. Such conditions hold if, for instance, the boundary CFT admits a parametrically large gap to the lightest spin-$\ell>2$ single-trace primary \cite{Heemskerk:2009pn,Afkhami-Jeddi:2016ntf,Camanho:2014apa,Fitzpatrick:2010zm} and, perhaps, a sparse spectrum of ``light" primaries.  Often, however, CFTs have no such gap, and one may adopt a simplified bulk gravitational description for the sake of expediency and tractability.

It is important to try to understand which features of boundary correlators in holographic theories are relatively insensitive to such simplifying assumptions, and which are not.  One recent result \cite{Fitzpatrick:2019zqz} along these lines, in the context of bulk gravity {\it minimally} coupled to a scalar field, is that in a certain lightcone limit, heavy-light correlators $\< \O_H\O_H\O_L\O_L\>$ -- i.e. scattering amplitudes of the light scalar field dual to $\O_L$ in a background created by the heavy operator $\O_H$ - are independent of all higher-curvature terms in the purely gravitational effective action at leading order in large central charge $C_T$.\foot{$C_T$ is defined as the norm of the stress tensor.} In CFT language, the non-trivial statement here pertains to the three-point coupling $\< \O_L\O_L [T^m]\>$, where $[T^m]$ are the $m$-trace stress tensor operators of lowest possible twist, $\tau\equiv \D-\ell = m(d-2)+\O(C_T^{-1})$. (We have suppressed the spin index on $[T^m]$.) For $m=2$, these ``minimal-twist'' composites are symmetric traceless primaries of spin-$\ell$, of the schematic form
\begin{align}
[TT]_{0,\ell}\approx T^{\mu_1\mu_2}\partial^{\mu_3}...\partial^{\mu_{\ell-2}}T^{\mu_{\ell-1}\mu_{\ell}},
\end{align}
where the ``$0$'' subscript denotes the condition of minimal twist. The claim of \cite{Fitzpatrick:2019zqz} is that if $\CO_L$ is dual to a minimally-coupled bulk scalar, then at large $C_T$, $\< \O_L\O_L [TT]_{0,\ell}\>$ depends only on the central charge $C_T$ and the dimension $\Delta_L$, and not on higher-derivative terms in the purely gravitational bulk action; likewise for the OPE coefficient with the $[T^m]$ operators. See \cite{Fitzpatrick:2019efk, Li:2019tpf, Kulaxizi:2019tkd,  Karlsson:2019dbd, Li:2019zba, Karlsson:2019txu, Li:2020dqm, Karlsson:2020ghx, Huang:2019fog, Huang:2020ycs, Parnachev:2020fna} for related recent works.

Understanding when such ``minimal-twist universality" holds or fails, or more precisely what additional data these minimal-twist OPE coefficients might depend on in a wider class of theories, is the motivation of the present work. From the point of view of bulk effective field theory, the restriction to the purely gravitational action is not parametrically controlled in known examples. In canonical instances of AdS/CFT with Einstein gravity coupled to low-spin matter in the bulk, there is abundant evidence that $\D_\gap$, the dimension of the lightest single-trace primary operator of spin $\ell>2$, gives the parametric dependence for higher-derivative gravitational interactions in AdS \cite{Camanho:2014apa,Afkhami-Jeddi:2016ntf}. Given that gravity-matter couplings appear in the derivative expansion at the same order as purely gravitational terms, one must contend with these couplings. This is what we will do here. We shall restrict ourselves to an investigation at leading order in large $C_T$, where the bulk description is classical, but we will not demand that the bulk scalar field is minimally coupled or that it is the only bulk matter field. 

The outcome is that certain bulk fields and interactions can indeed modify the minimal-twist OPE coefficients without modifying $C_T,\Delta_L$ or $\Delta_H$. In addition, we provide an efficient proof, not requiring explicit computation, that the minimal-twist OPE data is independent of the purely gravitational action, that is, of $n$-point graviton self-interactions.

The basic mechanism for the corrections is easy to explain with an example. Let us phrase this in terms of CFT. To access the minimal-twist OPE coefficient $\< \O_L\O_L [TT]_{0,\ell}\>$, we can study the four-point function $\<T \O_L\O_L T\>$: at leading-order in large $C_T$ (i.e. tree-level in AdS), the $[TT]_{0,\ell}$ operators are only exchanged in the $t$-channel, $\O_L\O_L \rar [TT]_{0,\ell} \rar TT$. Suppose that there exists a $\< TT\O_L\>$ coupling. This will contribute in the $s$-channel, $T \O_L \rar T \rar T \O_L$. By crossing symmetry, its presence will, barring kinematic cancellations, modify the $t$-channel exchange of $[TT]_{0,\ell}$ \cite{Komargodski:2012ek,Fitzpatrick:2012yx,Liu:2018jhs}. The explicit computations herein show that, indeed, such cancellations do not happen. Therefore, if $\< TT\O_L\>$ has the necessary scaling with $C_T$ to contribute to the leading-order correlator, $\< \O_L\O_L [TT]_{0,\ell}\>$ will pick up a dependence on that coupling. As may be familiar from tree-level AdS/CFT computations, the ``necessary scaling'' is
\be
\< TT\O_L\> \sim \sqrt{C_T}\,,
\ee
Moreover, as shown in \cite{Meltzer:2017rtf,Afkhami-Jeddi:2018own}, the $\< TT\O_L\>$ coupling (which carries a unique tensor structure) is, in CFTs with a large higher-spin gap, proportional to $\D_\gap^{-2}$. Therefore, this $\< TT\O_L\>$ coupling induces a perturbative $\D_\gap$-dependence of the minimal-twist OPE data. 

Let us elaborate on this, and our approach from the bulk perspective. The kinds of additional bulk terms we are interested in can be schematically written as 
\be
\label{phiaction}
S  \supset \int d^{D}x \sqrt{g} ~\Big( {\cal L}_{\rm grav} + {\cal L}_{\phi W^m} + {\cal L}_{\nabla^n \phi \nabla^n \phi^*  W^m}+ {\cal L}_{\phi W^m V}\Big)   
\ee
with
\bea
&&{\cal L}_{\rm grav}=R+ \Lambda + \sum_i \alpha_i  {\cal O} (R^2) + \sum_j \beta_j  {\cal O} (R^3) +\cdots  \label{type1} , \\
\label{type2}
&&{\cal L}_{\phi W^m}= \sum_{m} a_{m} \phi {W}^m  ,  \\   
\label{type3}
&&{\cal L}_{\nabla^n \phi \nabla^n \phi^*   W^m} = \sum_{n,m}   b_{n,m} (\nabla^n \phi \nabla^n \phi^* ) W^m  , \\
\label{type4}
&&{\cal L}_{\phi W^m V} = \sum_{m, \ell } c_{m, \ell } \phi W^m V_\ell,     
\eea  
The notation is as follows: $W$ is the Weyl curvature; $\nabla$ is the covariant derivative; the superscripts indicate powers (not indices); $a, b$ and $c$ are bookkeeping constants; each term generally stands for multiple different terms where the indices (which have been suppressed above) may be contracted differently; and the field $V_\ell$ indicates a bulk field with spin $\ell\leq 2$, where the bound follows from imposing Einstein gravity at low energies. We will call \eqref{type1} the purely gravitational part of the action. For each of \eqref{type2}$-$\eqref{type4}, in this work we will study the effect of one simplest representative interaction; the generalization of our methods to other operators is straightforward in principle.\foot{We have chosen to write these interactions in terms of the Weyl tensor instead of the Riemann tensor because Riemann is non-vanishing in pure AdS, and therefore interactions written in terms of Riemann typically contain some contributions that can be absorbed by shifts in bare parameters for lower-order couplings.} The $\< TT\O_L\>$ coupling described above is dual to the Weyl-squared coupling,
\be
\CL_{\phi W^2} \sim 
 \phi W^{\mu\nu\rho\sigma}W_{\mu\nu\rho\sigma}.
\ee

We will demonstrate that the minimal-twist OPE data is independent of $\CL_{\rm grav}$ and show how each of the other classes of vertices corrects this data.  
For example, ${\cal L}_{\phi W^m}$ interactions can contribute at large $C_T$ only if the self-interactions of the probe field $\phi$ are ``gravitationally suppressed,'' i.e. suppressed by appropriate powers of $C_T$. This form of large-$C_T$ factorization is not known to be required by any first principles argument. On the other hand, it is obeyed in UV complete constructions of AdS vacua. Some familiar cases of scalars in large-$C_T$ CFTs with this scaling include the dilaton (either in type II string theory or in more general KK reductions); all known reliable AdS$\times \mathcal{M}$ compactifications of low-energy string or M-theory, which have $L_{\mathcal{M}}\sim L_{\rm AdS}$; and SCFTs whose stress tensor multiplet contains a scalar primary (see section \ref{s22}). In what follows we sometimes refer to this large-$C_T$ scaling as ``gravitational scaling''.

On the other hand, the bulk interactions $\nabla^n \phi \nabla^n \phi^* W^m$ do not require a specific large-$C_T$ scaling but they can also affect the minimal-twist sector; this type of correction may be generated by integrating out a spin-$\ell$ $\geq 2$ bulk matter at tree level, or any spin at loop level.\footnote{The ${\phi W^m V}$ interactions do not require a specific scaling either, and they can affect the minimal-twist OPEs when $V_\ell$ has $\ell=2$.}

A more detailed accounting of the paper is as follows.  

In section \ref{sec:bulkinteractions}, we elaborate on the different forms of interactions, how they might scale at large $C_T$, and the types of AdS/CFT dual pairs for which this scaling holds. 

In sections \ref{s3} and \ref{sec:explicitcomputations2}, we compute the corrections to the OPE coefficients of minimal-twist multi-$T$ operators in two related scenarios. 

First, in section \ref{s3}, we extract $\< \O_L\O_L [TT]_{0,\ell}\>$ from the four-point function $\< T\O_L\O_L T\>$, where the corrections come from spin-two exchange. More specifically, we compute the correction to the OPE coefficient at $\ell\gg1$ using the spinning lightcone bootstrap for spin-two exchange in AdS. This exemplifies the effect of the $\CL_{\phi W^2}$ and $\CL_{\phi W V}$ vertices shown above. This approach also gives an efficient argument for why $\<\mathcal{O}\mathcal{O}[TT]_{0,\ell}\>$ (as well as all the minimal-twist $\< \CO \CO [T^m]\>$ OPE coefficients) are independent of the purely gravitational action $\CL_{\rm grav}$  (see section \ref{gravind}).

Then in section \ref{sec:explicitcomputations2}, we switch gears and study the effect of quartic bulk interactions, of the type ${\cal L}_{\nabla^n \phi \nabla^n \phi^*   W^2}$ shown above, on the four-point function $\< \O_H\O_H\O_L\O_L\>$. This is done by solving bulk equations of motion. We fix $d=4$ for simplicity. This method allows us to extract $\< \O_L\O_L [T^m]\>$ where here, in a slight abuse of notation, $[T^m]$ are the minimal-twist operators of spin $\ell=2m$, 
\be
[T^m] \equiv T_{\mu_1\nu_1}T_{\mu_2\nu_2}\ldots T_{\mu_m\nu_m}~.
\ee
We find for the first several values of $m$ that indeed, these minimal-twist OPE coefficients are modified by ${\cal L}_{\nabla^n \phi \nabla^n \phi^*   W^2}$.

Finally, we conclude in section \ref{sec:discussion}. Appendix \ref{app:spinningdetails} gives further details on the lightcone bootstrap for spinning operators and Appendix \ref{app:Loops} explains how our results are consistent with earlier work on universality and the inversion formula. 


\section{Higher-curvature interactions at large $C_T$}
\label{sec:bulkinteractions}

In this section, we make some general remarks on the higher-derivative corrections 
to a bulk gravitational action, focusing on the large central charge limit.
In particular, we shall point out several bulk coupling terms which were not included 
in the earlier analysis \cite{Fitzpatrick:2019zqz} and argue that some of these additional couplings 
can {\it a priori} contribute to the lowest-twist stress tensor sector.  We confirm this in detailed computations in later sections.

\subsection{Neutral vs charged scalar}

We begin with the purely gravitational part of the action:
\be
S_{\rm grav} \sim C_T \int d^D x \sqrt{g} \Big( R+ \Lambda+\sum_i  \alpha_i {\cal O} (R^2)_i+\sum_j \beta_j {\cal O} (R^3)_j +\sum_k \gamma_k{\cal O} (R^4)_k+\cdots \Big) 
\label{eq:gravaction}
\ee
where ${\cal O} (R^{n})$ denotes all possible invariants constructed out of $n$ Riemann tensors.
The coefficients $\alpha_i, \beta_j, \dots$  in  $S_{\rm grav}$ are assumed only to be finite in the $C_T\to \infty$ limit; in particular, we make {\it no} assumption about these coefficients being suppressed by an EFT gap scale, such as $\D_\gap$.   
We also introduce a light bulk scalar field $\phi$.\footnote{By ``light", we mean fields whose masses are finite at $C_T \rightarrow \infty$, whereas ``heavy" will indicate masses that grow linearly with $C_T$. }  
At a minimum, its coupling to gravity includes the standard kinetic term, $S_{\rm kin} \sim \int d^D x \sqrt{g} g^{\mu\nu} \partial_\mu \phi \partial_\nu \phi$. 

If $\phi$ is neutral under all symmetries, then even in the absence of additional fields in the bulk, the bulk action may contain terms that are linear in $\phi$. The simplest such term that does not produce a tadpole for $\phi$ in the pure AdS background is 
\be
\CL_{\phi W^2} \sim 
 \phi W^{\mu\nu\rho\sigma}W_{\mu\nu\rho\sigma},
 \label{eq:phiC2}
\ee
where $W_{\mu\nu\rho\sigma}$ is the Weyl tensor.   
This bulk interaction reflects the presence of a $\< TT\CO_L\>$ three-point function in the boundary CFT. It  is a special case of an infinite family of interactions  (\ref{type2}).  Below, we will consider in detail the effect of (\ref{eq:phiC2}) on the OPE coefficients $\< \O_L\O_L [TT]_{0,\ell}\>$ and comment on more general cases.

By contrast, if $\phi$ is charged, (\ref{eq:phiC2}) is forbidden.  Instead, we can either form neutral combinations out of its products  or introduce additional charged fields, $V$, into the theory. 
For example, we can consider 
\be
\label{eq:ddphi2C2}
\CL_{\nabla^2 \phi \nabla^2 \phi^* W^2} \sim \nabla_\mu \nabla^\nu \phi \nabla_\rho \nabla^\sigma \phi^* W^{\alpha \mu \beta \rho}W_{\alpha \nu \beta \sigma},
\ee
or
 \be
 \CL_{\phi W V} \sim \nabla^{\nu} \nabla^\sigma \phi W_{\mu\nu\rho\sigma} V^{\mu \rho} ,
 \label{eq:phiCV}
 \ee
 where $V$ is a charged symmetric traceless field of spin $\ell=2$.  These are special cases of the families (\ref{type3}) and (\ref{type4}), respectively. The exact structure of the index contractions here is not important for the moment. Note that the two options (\ref{eq:ddphi2C2}) and (\ref{eq:phiCV}) are not so different: in the limit where the additional charged fields $V^{\mu\rho}$ become heavy, integrating them out leads to additional local interactions of $\phi$ and $h_{\mu\nu}$ made from neutral combinations of the scalar field. The main physical difference between these two cases is, therefore, just the difference between the effect of local interactions and nonlocal exchange; in practice, we will use rather different methods to compute their effects on the minimal-twist multi-$T$ OPE coefficients.

\subsection{Large $C_T$ and gravitational scaling}\label{s22}

In order to determine whether the interactions above contribute to the minimal-twist OPE coefficients in the $C_T\rightarrow \infty$ limit, we need to consider how the coefficients of these interactions scale with $C_T$. If they grow too slowly, then the inverse factors of $C_T \sim G_N^{-1}$ from gravitational suppression will cause their effect to vanish. By contrast, if they grow too quickly, they can destabilize the classical limit of the gravitational description. 

 Perhaps the simplest way to think about this is to imagine that, as is often the case, the large $C_T$ in the CFT description arises from the large dimension -- call it $N$ -- of a symmetry group, with the bulk fields being composite singlets.  Then all disconnected $n$-point functions of the stress tensor $T$ are naturally proportional to 
\be
C_T \sim N \ , 
\ee 
and the large $N$ bulk action automatically has a factor of $C_T$ out front as in (\ref{eq:gravaction}), and a classical limit emerges.  Now imagine that the CFT dual $\CO$ of the light field $\phi$ is also a composite of a large number $M$ of constituents $q_i$, e.g. 
\be
\CO \sim \sum_{i=1}^M \bar{q}_i q_i.
\ee
If $M$ is large, then $\phi$ will also have a weakly-coupled bulk description where the kinetic term and bulk interactions involving $\phi$ are proportional to $M$.\foot{Note that there is no meaningful limit where $M \gg N$.  From the CFT point of view, the stress tensor couples to all the degrees of freedom in the theory, so $N$ should be at least as large as $M$. From the bulk point of view, if one set $M \gg N$ by hand, $\phi$ loops would simply renormalize $N$ back up to be $\sim M$.}  If $M$ and $N$ are the same parametrically, then we can write the action for all the bulk terms involving $\phi$ and $h_{\mu\nu}$ with a single factor of $C_T$ out front:
\be
S \sim S_{\rm grav} + C_T \int d^D x \sqrt{g} \left( (\nabla \phi)^2 + a \phi W^2 + a' (\nabla \nabla \phi)^2 W^2 + \dots \right), ~~~ (M \sim N).  
\ee
In this case, the interaction terms involving $\phi$ affect the large $C_T$ classical gravity equations of motion.  If these interactions affect the minimal-twist multi-$T$ OPE coefficients -- which, as we show in the next section, they do -- the effect will therefore survive at $C_T\rar\i$. 

By contrast, if $M$ is parametrically smaller than $N$, then we expect the coefficient in front of the part of the action involving $\phi$ to be at most $M$,
\be
S \sim S_{\rm grav} + M \int d^D x \sqrt{g} \left( (\nabla \phi)^2 + a \phi W^2 + a' (\nabla\nabla \phi)^2 W^2 + \dots \right),~~~(M \ll N)  
\ee
Then $\phi$ will not affect the classical gravity equations of motion. 
 In terms of the method for computing these OPE coefficients used in \cite{Fitzpatrick:2019zqz}, when $M\ll N$ we can first solve for the metric created by the heavy operator while ignoring $\phi$, and then solve for the bulk $\phi$ two-point function in that background. The $\phi$ equations of motion take the schematic form
\be
\nabla^2 \phi \sim a W^2 + a' \nabla \nabla\nabla\nabla \phi W^2 + \dots, 
\ee
The key difference between the case $M \sim N$ and $M \ll N$ is as follows. If $M \ll N$, then we do not also have to consider contributions where two factors of $h_{\mu\nu}$ contract with each other in a diagram with an internal graviton propagator: due to the $C_T^{-1} \sim N^{-1}$ suppression in the graviton propagator,  such terms will be suppressed by $M/N \ll 1$ in this case and therefore subleading. 

It is also useful to state these considerations in a convention where we canonically normalize $\phi \rightarrow M^{-1/2} \phi$.  In that case, we would write the $\phi$-field bulk action schematically as
\bea
\label{newform}
S \sim \int d^D x \sqrt{g} \left( (\nabla \phi)^2 + a M^{1/2} \phi W^2 + a'(\nabla \nabla \phi)^2 W^2 + \frac{\phi^n}{M^{\frac{n-2}{2}}} + \dots\right) ,  ~~  n \in  \mathbb{N}, ~ n >2 \ . 
\eea
In this language, when $M \sim N$, 
\begin{center}
\begin{quote}
{~~~~~~~~~ \it all $\phi^n$ with $n\geq 3$ interactions must be gravitationally suppressed.}
\end{quote}  
\end{center}
By ``gravitationally suppressed'' we mean suppressed by powers of $C_T\sim N$. 

\begin{figure}[t!]
\begin{center}
\includegraphics[width=0.25\textwidth]{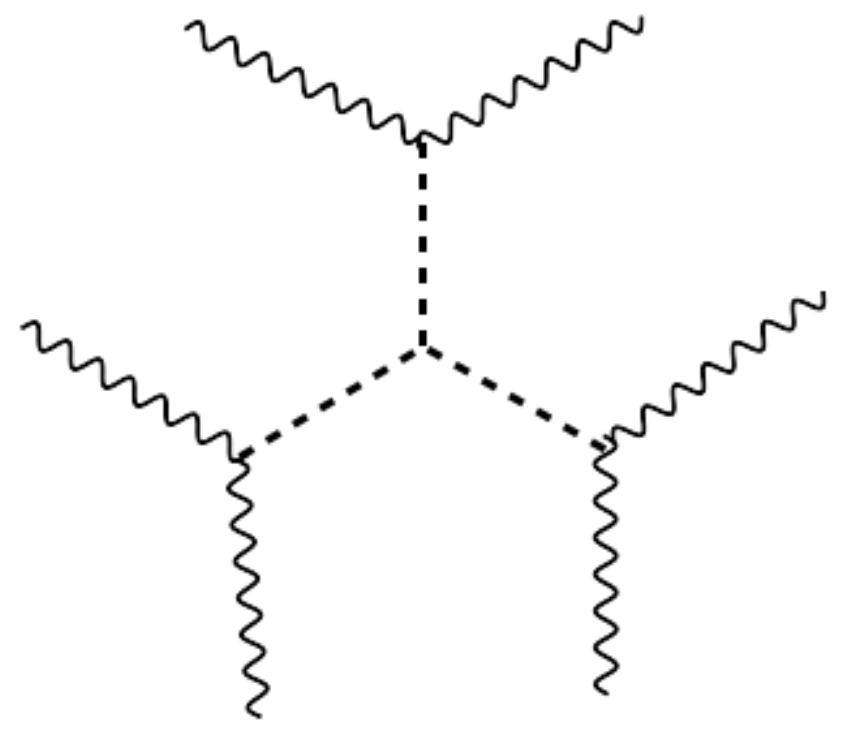} \hspace{1in}
\includegraphics[width=0.22\textwidth]{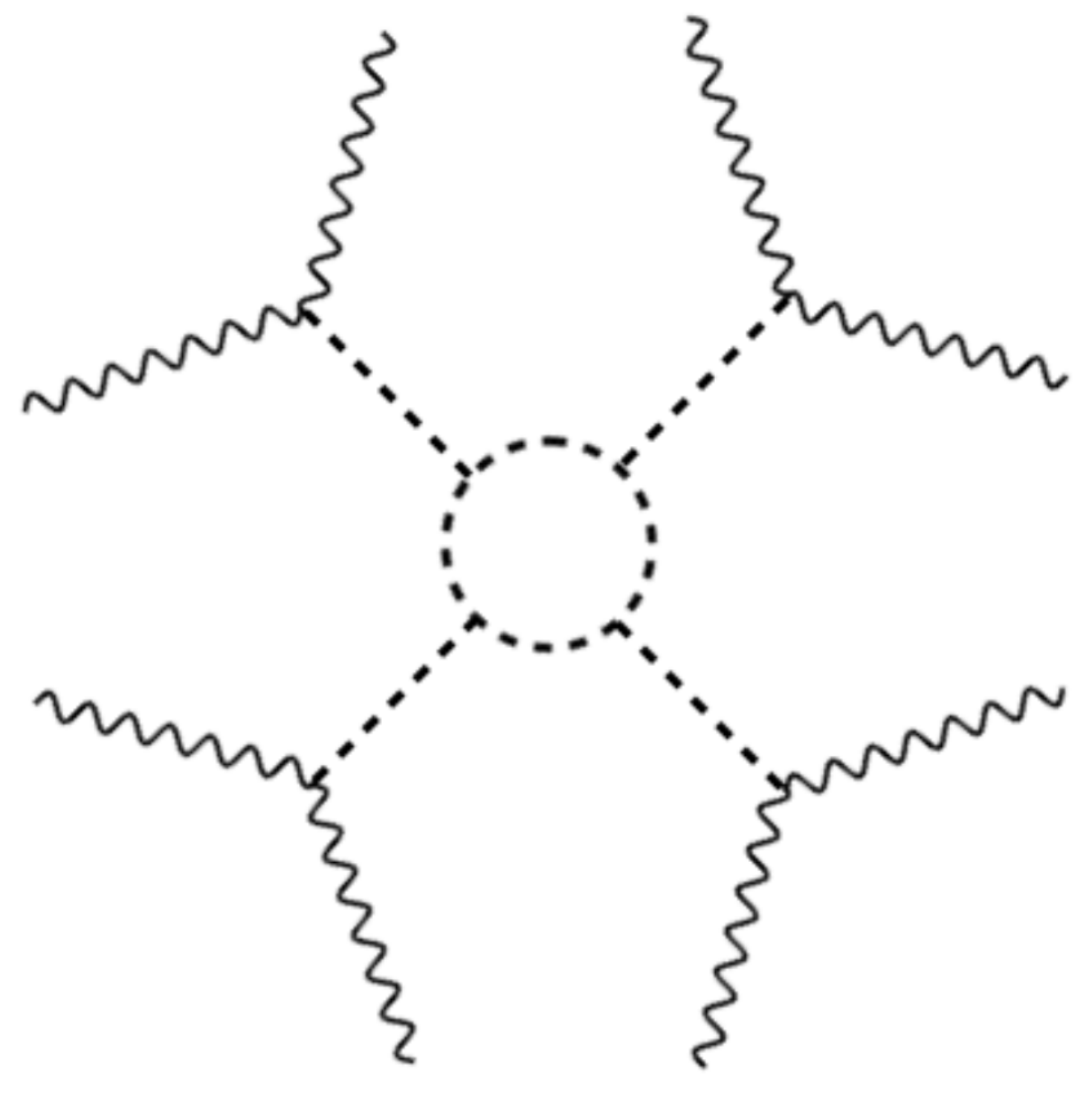}    
\caption{Generating higher-curvature terms by integrating out $\phi$ with a $g \phi^3$ interaction.   
Dashed lines are $\phi$ propagators, and external gravity lines represent insertions of the  Weyl tensor.}
\label{fig:generatingR6}
\end{center}
\end{figure}

One can derive this fact in the opposite direction as well, by putting in a $(C_T)^{\alpha} \phi W^2$ interaction and a $\phi^n$ interaction without any $1/C_T$ suppression, and seeing that $\phi$ loops generate a gravitational action that does not have the form (\ref{eq:gravaction}), and does not have a classical limit.  In figure \ref{fig:generatingR6}, we show examples generating $g (C_T)^{3\alpha} W^6$ and $g^4 (C_T)^{4\alpha} W^8$ in the presence of a $g \phi^3$ interaction.  By adding more $g \phi^3$ insertions on the $\phi$ loop in figure~\ref{fig:generatingR6}, we can generate $g^{n} (C_T)^{n \alpha} W^{4n}$ for any $n$.  We therefore see that if $g$ does not have any $C_T$ suppression, then any $\alpha > 0$ leads to terms in the gravitational action that are larger than that allowed in (\ref{eq:gravaction}). 
Conversely, if $\alpha = \frac{1}{2}$, as adopted in \eqref{newform}, then we must take $g \lesssim \CO(\frac{1}{\sqrt{C_T}})$.

Summarizing, in order to affect the minimal-twist OPE coefficients via the coupling $\phi W^2$, the corresponding scalar must have self-interactions suppressed by the powers of $C_T$ just described. As noted earlier, this scaling is not known to be required by any holographic consistency condition. On the other hand, this $C_T$-dependence is obeyed by scalars in controlled constructions of AdS vacua. This includes:

\begin{enumerate}[font=\bfseries]

\item The dilaton, either in type II string theory or in more general KK reductions, whose interaction are suppressed by gravity.

\item All known reliable AdS$\times \mathcal{M}$ compactifications of low-energy string or M-theory which have $L_{\mathcal{M}}\sim L_{\rm AdS}$, and hence KK scalars with masses of order the AdS scale. These solutions also always contain KK towers of massive spin-two operators, descending from the graviton, and hence can furnish all types of vertices given in the previous subsection. 

\item Any SCFT whose stress tensor multiplet contains a neutral scalar field. This follows from the fact that $\< TTT\>$ may be generated from $\< TT\O\>$ via the action of supersymmetry. This includes, in particular, 3d $\mathcal{N}=4$ and 4d $\mathcal{N}=2$ SCFTs (e.g. \cite{Cordova:2016emh}), in which $\O$ is the bottom component of the supermultiplet.

\end{enumerate}

\noindent We will consider the effect of the interactions obeying gravitational scaling on the minimal-twist OPE coefficients in the following section.


\section{Exchange interactions}
\label{s3}

We now explicitly compute the corrections to the minimal-twist OPE coefficients $\<\mathcal{O}\mathcal{O}[TT]_{0,\ell}\>$ from interactions of the forms (\ref{eq:phiC2}) and (\ref{eq:phiCV}) discussed in the previous section. Our strategy will be to extract the desired OPE coefficients from the four-point function $\< T\O\O T \>$. This correlator contains exchange diagrams with massless or massive spin-two exchange, allowed by the interactions (\ref{eq:phiC2}) and (\ref{eq:phiCV}), and we show that they do indeed contribute to the minimal-twist OPE coefficients.

This point of view also provides a compact argument, given in subsection \ref{gravind}, for why the OPE coefficients $\<\mathcal{O}\mathcal{O}[T^m]\>$ are independent of the purely gravitational sector.

As before, we will denote the boundary scalar by $\mathcal{O}$ and its bulk dual by $\phi$. In this section (except subsection \ref{gravind}), $\phi$ is assumed to be a neutral field with gravitationally-suppressed interactions.

\subsection{Witten diagrams}
\label{sec:WittenDiagramsExchange}

\begin{figure}
\begin{center}
\includegraphics[scale=.27]{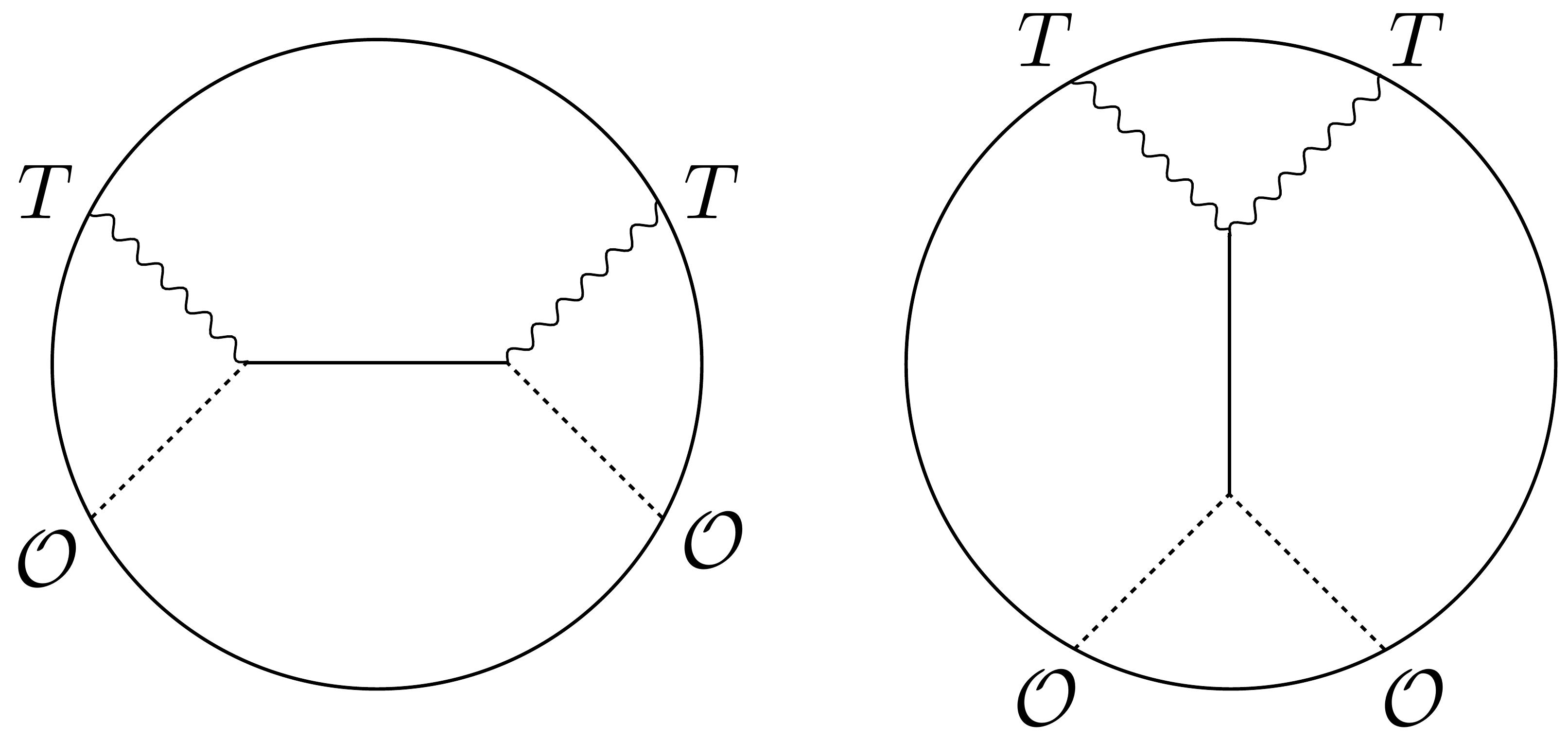}
\caption{$s$ and $t$-channel exchange diagrams for $\<T\mathcal{O} \mathcal{O}  T\>$, respectively.}
\label{fig:Witten_Diagrams_TffT}
\end{center}
\end{figure}

In order to classify which interactions can affect the coupling $\<\mathcal{O}\mathcal{O}[TT]_{0,\ell}\>$ it is simplest to think in terms of Witten diagrams.  We are interested in the leading large $C_{T}$ correction, so only tree diagrams for $\<T\mathcal{O}\mathcal{O} T\>$ are relevant since both $T$ and $\O$ have $\D\ll C_T$.  
There are two types of exchange diagrams we have to study, as shown in figure \ref{fig:Witten_Diagrams_TffT}. 

First, it is straightforward to see that no diagrams in the $t$-channel, $\mathcal{O} \mathcal{O} \rightarrow \chi \rightarrow TT$, can affect the minimal-twist OPE coefficients. This follows from known results. First, recall that the minimal-twist double-trace stress tensors operators $[TT]_{0,\ell}$ start at $\ell=4$. Next, it is known \cite{Sleight:2017fpc} that in a direct-channel decomposition, the spin of the double-trace operators for which OPE data is generated is bounded by the spin of the exchanged single-trace operator. In the $t$-channel, only symmetric traceless fields of spin-$\ell$ may be exchanged. Since we are looking at a theory of Einstein gravity + matter in the bulk, all elementary bulk fields have \cite{Camanho:2014apa,Maldacena:2015waa}
\be\label{spin2}
\ell \leq 2 \ .
\ee 
This concludes the proof. 

Thus, turning to the $s$-channel,  $T\mathcal{O}  \rightarrow  \mathcal{\chi}  \rightarrow  \mathcal{O} T$, we must classify which fields $\chi$, dual to single-trace operators, can be exchanged.  One universal set of operators are the symmetric, traceless operators, i.e. operators that transform in the representation $[\ell] \equiv [\ell,0,\ldots ,0]$ of the Lorentz group. Here $[\ell_1,\ell_2,...,\ell_n]$ gives the number of boxes in each row of the Young tableaux and $\ell_1\geq\ell_2\geq...\geq\ell_n$. In $d=3$ these are the only operators which can appear. In $d\geq 4$ we can also have the mixed symmetry operators $[\ell,1]$ and $[\ell,2]$\cite{Costa:2014rya}. 

The Lorentzian inversion formula tells us that inverting a single operator in one channel gives corrections to the double-trace operators in the crossed-channel, for all spin and twist \cite{Caron-Huot:2017vep}. Therefore, for $\<T \mathcal{O} \mathcal{O} T\>$ we expect that single-trace exchange in the $s$-channel corrects the couplings to the $t$-channel double-traces, which here are $[TT]_{n,\ell}$ and $[\mathcal{O} \mathcal{O}]_{n,\ell}$. In general dimensions $d$, the OPE coefficients $\<\mathcal{O} \mathcal{O} [TT]_{0,\ell}\>$ should depend on couplings to the three Lorentz tensor structures, $\<\mathcal{O} T \chi_{\Delta,[\ell]}\>$, $\<\mathcal{O}  T \chi_{\Delta,[\ell,1]}\>$, and $\<\mathcal{O}  T \chi_{\Delta,[\ell,2]}\>$. It will be sufficient for our purposes to focus on the symmetric traceless operators. Then to determine if a bulk Witten diagram is allowed or not, we need to classify the allowed three-point functions  $\< T\mathcal{O} \chi_{\Delta,[\ell]}\>$ for $\ell=0,1,2$, and impose conservation. 

To perform computations it is simpler to work in a $d+2$-dimensional embedding space with signature $(1,d+1)$ \cite{Costa:2011mg,Costa:2011dw}.\footnote{Here we are working in $d$ Euclidean dimensions, but later we will Wick rotate to Lorentzian signature.} We will write CFT three-point functions in terms of the $d+2$-dimensional, null position and polarization vectors, $P$ and $Z$ and choose the following basis of tensor structures, 
\begin{align}
k_{ijk}=\frac{P_{ij}Z_{i}\cdot P_{k}-P_{ik}Z_{i}\cdot P_{j}}{\sqrt{P_{ij}P_{jk}P_{ik}}} \ , ~~~~~~~ m_{ij}=Z_i\cdot Z_{j}-\frac{Z_{i}\cdot P_{j} Z_{j}\cdot P_{i}}{P_{i}\cdot P_{j}}\ ,
\end{align}
where $P_{ij}=-2 P_{i}\cdot P_{j}$. The three-point function takes the form
\begin{align}
\<T \mathcal{O}  \chi_{\Delta,\ell}\>=\frac{1}{P_{12}^{\Delta_{T\mathcal{O}  \chi}}P_{23}^{\Delta_{\mathcal{O}  \chi T}}P_{13}^{\Delta_{T \chi\mathcal{O} }}}\sum\limits_{i=0}^{2}\lambda^{(i)}_{T\mathcal{O} \chi}k^{2-i}_{1}k_{3}^{\ell-i}m_{13}^i
\end{align}
where $k_{1}=k_{123}$, $k_{3}=k_{312}$ and $\Delta_{123}=\Delta_1+\Delta_2-\Delta_3$. 
We next need to impose conservation for the stress tensor to see which three-point functions are allowed. In embedding space, the conservation operator is
\begin{align}
D_{i}=\frac{\partial}{\partial P_{i,M}}\left[ \left(\frac{d}{2}-1+Z\cdot \frac{\partial}{\partial Z}\right)\frac{\partial}{\partial Z^{M}}-\frac{1}{2}Z_{M}\frac{\partial^{2}}{\partial Z\cdot \partial Z}\right].
\end{align}
We can impose the condition $D_1\<T \mathcal{O} \chi_{\Delta,\ell}\>=0$ directly in embedding space.

The simplest case is when $\ell=0$, in which case $\chi_{\Delta,\ell=0}$ is a scalar. There is a unique allowed tensor structure and imposing conservation implies $\Delta=\Delta_{\mathcal{O} }$. The stress tensor Ward identity at coincident points implies we must have $\chi=\mathcal{O}$ itself and fixes the OPE coefficient to
\begin{align}\label{ward}
\lambda_{\mathcal{O}  \mathcal{O}  T}=-\frac{\Delta_{\mathcal{O} }d}{\sqrt{C_{T}}(d-1)S_{d-1}}
\end{align}
where $S_{d-1}$ is the area of a $d-1$ dimensional sphere. In other words, the only scalar which can appear in the $s$-channel is $\mathcal{O} $ itself and the coupling is fixed to the minimally coupled answer. 

Next, we consider the exchange of an $\ell=1$ operator, $V$. Imposing conservation of $T$ relates the two tensor structures as \cite{Meltzer:2017rtf},
\begin{align}
\lambda_{T\mathcal{O}  V}^{(0)}=\frac{1}{2}\lambda_{T\mathcal{O}  V}^{(1)}(d(\Delta_{\mathcal{O} }-\Delta_{V})+2) \ , ~~~~ \Delta_{\mathcal{O} }=\Delta_{V}\pm 1 \ .
\end{align}
If we furthermore impose that $V$ is conserved, $\Delta_{V}=d-1$, then we must set $\Delta_{\mathcal{O} }=d-2$. 
This is a somewhat strange cubic coupling; the conservation implies that it is only allowed if the dimensions of $\mathcal{O} $ and $V$ are correlated. 
We are not aware of such an allowed interaction in a holographic CFT and we shall not consider this case any further.\footnote{This coupling can be non-zero in a free field theory \cite{Giombi:2011rz}.}

Finally, we consider the exchange of an $\ell=2$ operator, $M$. This could be, for example, a massive KK mode of a higher-dimensional graviton. Imposing conservation gives a unique tensor structure, where the three {\it a priori} independent structures are related as
\begin{align}
\lambda^{(0)}_{T\mathcal{O}  M}&=\frac{(d-2)  (d+\Delta_{\mathcal{O} }-\Delta_{M}) (d+\Delta_{\mathcal{O} }-\Delta_{M}+2)}{d \left((\Delta_{\mathcal{O} }-\Delta_{M})^2-2\right)-(\Delta_{\mathcal{O} }-\Delta_{M})^2}\lambda^{(2)}_{T \mathcal{O}  M} \ ,
\\
\lambda^{(1)}_{T\mathcal{O}  M}&=\frac{2  (d+\Delta_{\mathcal{O} }-\Delta_{M}) ((d-1) \Delta_{\mathcal{O} }-d \Delta_{M}+\Delta_{M}-2)}{d \left((\Delta_{\mathcal{O} }-\Delta_{M})^2-2\right)-(\Delta_{\mathcal{O} }-\Delta_{M})^2}\lambda^{(2)}_{T\mathcal{O}  M} \ .  \label{eq:ConsTphiV}
\end{align}
The three-point function $\<T\mathcal{O}  T\>$ is recovered by setting $\Delta_{M}=d$.
We see that spin-two exchange in the $s$-channel depends on a free parameter; thus, absent kinematic cancellations, one would expect this to correct the minimal-twist universality. The next subsection confirms this expectation. 

\subsubsection{Minimal-twist data is independent of $\CL_{\rm grav}$}\label{gravind}

\begin{figure}[t!]
\begin{center}
\includegraphics[width=0.65\textwidth]{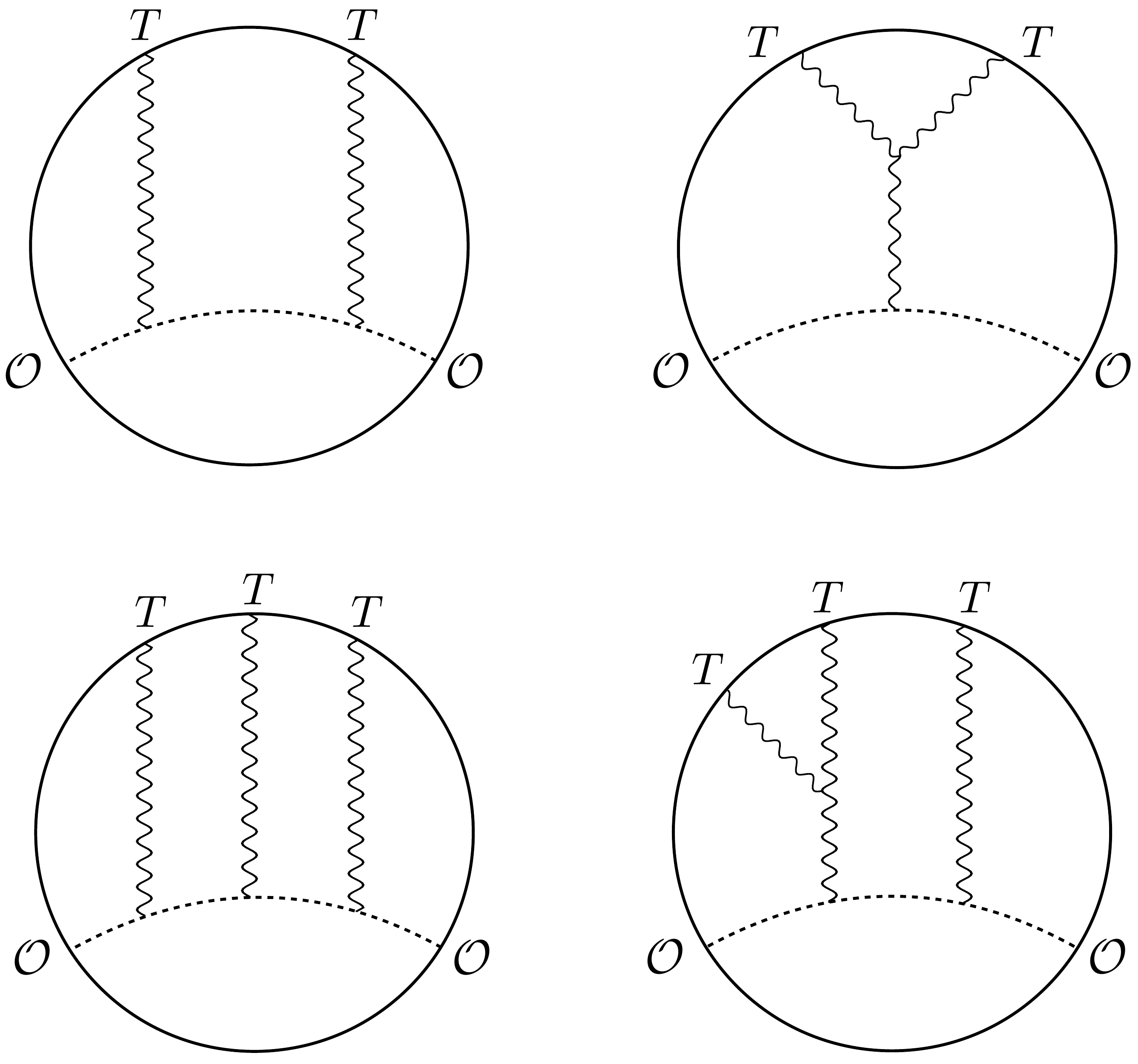}
\caption{Examples of tree-level diagrams for  the $\< {\cal O} {\cal O} [T^m]\>$ OPE coefficients in the case of a minimally-coupled scalar.  The left two diagrams contribute to the minimal-twist $[T^m]$ OPE data, whereas the right two diagrams do not.}
\label{treegravproof}
\end{center}
\end{figure}

Before moving on, we pause to note that the above perspective gives a direct explanation of why the OPE coefficients $\<\mathcal{O}\mathcal{O}[TT]_{0,\ell}\>$ depend on the purely gravitational action, $S_{\rm grav}$, only through the central charge $C_T$. Higher-curvature terms in the gravitational action only modify $\< T\O\O T\>$ at tree-level via the cubic coupling $\< TTT\>$, and $\< TTT\>$ only contributes to $\< T\O\O T\>$ in the $t$-channel, $\O\O \rar T \rar TT$, as depicted in the upper right diagram of figure~\ref{treegravproof}. This is the same channel in which the minimal-twist double-traces appear. But as recalled earlier, a spin-$j$ exchange does not affect direct-channel OPE data for double-trace operators of spin greater than $j$. Therefore, $T$ exchange cannot affect the couplings to the minimal-twist trajectory $[TT]_{0,\ell}$, which starts at $\ell=4$. 

This conclusion also gives a concise way to characterize the OPE coefficients $\<\mathcal{O}\mathcal{O}[TT]_{0,\ell}\>$: they are determined solely by the crossing transformation of the $T\O\rar \O \rar T\O$ exchange. This exchange is fixed by the conformal Ward identity as discussed above \eqref{ward}. Applying crossing to this exchange generates a $6j$ symbol \cite{Hogervorst:2017sfd,Liu:2018jhs,Sleight:2018ryu} whose residue at $\D=2d+\ell$ is the product $\<\mathcal{O}\mathcal{O}[TT]_{0,\ell}\>\<TT[TT]_{0,\ell}\>$, and the factor $\<TT[TT]_{0,\ell}\>$ is determined by stress tensor mean field theory (MFT).\foot{Techniques for computing stress tensor MFT OPE coefficients and explicit expressions in $d=3$ are given in \cite{Karateev:2018oml}.} In this way, $\<\mathcal{O}\mathcal{O}[TT]_{0,\ell}\>$ is determined by conformal kinematics alone.

Moreover, it easily generalizes to the case of minimal-twist operators $[T^m]$ with any $m$.  To see this for $m=3$, note that the tree-level diagrams for the $\< \O \O [T^3]\>$ OPE coefficient are either of the form of the lower left diagram in figure~\ref{treegravproof}, where no graviton self-interaction $n$-point vertices are present, or of the form of the lower right diagram, where such vertices are present.  The former case clearly is insensitive to cubic and higher graviton couplings.  The latter case does depend on such couplings, but does not affect the minimal-twist OPE coefficients for the following reason: the minimal-twist $[T^m]$ operators are also minimal twist when looking only at any subset of $T$s in the composite, so by looking at the subdiagram with two external $T$s connected by a $TTT$ vertex (equivalent to the upper right diagram), we see that the OPE coefficient vanishes by the same argument just given for the minimal-twist $[T^2]$ OPE coefficients.  

We can make one further generalization.  Suppose that we add any finite number of (possibly massive) scalar and spin-one bulk fields to the theory.  Assume that $\CO$ is charged under some symmetry, under which these new fields are neutral, so that the $\CO$ line is unbroken in any bulk diagram.  Then all new allowed tree-level diagrams that correct the $\< \O \O [T^m]\>$ OPE coefficients can be obtained by diagrams such as those shown in figure~\ref{treegravproof}, with some internal graviton lines replaced by scalar or gauge field lines.  Since the new fields have strictly lower spin than $T$, the previous proof goes through as before, and the minimal-twist $\< \O \O [T^m]\>$ OPE coefficients are unmodified.

\subsection{Lightcone bootstrap}
\label{sec:LightconeBootstrap}

With the results of section \ref{sec:WittenDiagramsExchange} in place, we now show explicitly that the minimal-twist OPE coefficients depend on the coupling $\lambda_{T\mathcal{O} M}$ where $M$ is a generic, non-conserved spin-two operator. We will use the lightcone bootstrap to extract the large-spin asymptotics for the spinning OPE data \cite{Komargodski:2012ek,Fitzpatrick:2012yx,Li:2015itl,Hofman:2016awc}, showing that it is nonzero. This will be sufficient to prove that universality is violated by the exchange of spin-two fields.\foot{To obtain the OPE coefficients at finite spin, one option would be to directly use the inversion formula for spinning operators, which can be derived through weight-shifting operators \cite{Karateev:2017jgd,Kravchuk:2018htv}.}

Our goal is to solve the crossing equation:
\begin{align}
 \label{eq:CrossingSpin2Full}
\mathcal{W}^{(s)}_{M}(x_i)=\sum\limits_{\ell=0}^{\infty}\lambda_{TT[TT]_{n,\ell}}\lambda_{\mathcal{O}\mathcal{O}[TT]_{n,\ell}}g^{T\mathcal{O}\mathcal{O}T}_{[TT]_{n,\ell}}(x_i)+\big([TT]_{n,\ell}\rightarrow [\mathcal{O}\mathcal{O}]_{n,\ell}\big) \ .
\end{align}
where $\mathcal{W}^{(s)}_{M}(x_i)$ is the $s$-channel exchange diagram for $M$ discussed in section \ref{sec:WittenDiagramsExchange}, and $g_{\Delta,\ell}(x_i)$ are the conformal blocks. Expanding the left-hand side of (\ref{eq:CrossingSpin2Full}) in $s$-channel conformal blocks, one finds blocks for the single-trace operator $M$ and the double-traces $[T\mathcal{O}]_{n,\ell}$. These double-trace operators do not affect the large spin asymptotics on the right-hand side of  (\ref{eq:CrossingSpin2Full}), so we can drop them.\footnote{In the language of the inversion formula, they have a vanishing double-discontinuity \cite{Caron-Huot:2017vep} or equivalently in the lightcone bootstrap, are Casimir-regular \cite{Simmons-Duffin:2016wlq}.} 
Therefore, we are left with the crossing equation:
\begin{align}\label{crosseq}
g^{T\mathcal{O}\mathcal{O} T}_{M}(x_i)\approx \sum\limits_{\ell=0}^{\infty}\lambda_{TT[TT]_{n,\ell}}\lambda_{\mathcal{O}\mathcal{O}[TT]_{n,\ell}}g^{T\mathcal{O}\mathcal{O} T}_{[TT]_{n,\ell}}(x_i)+\big([TT]_{n,\ell}\rightarrow [\mathcal{O}\mathcal{O}]_{n,\ell}\big).  
\end{align}
The $\approx$ is to emphasize that we will only be solving this equation in the lightcone limit and determining the large-spin asymptotics of the right-hand side. 

We will work with the $d$-dimensional metric,
\begin{align}
ds^{2}&=dx^+dx^-+\delta_{ij}dx^{i}dx^{i}, \qquad x^{\pm}=x^1\pm x^0,
\end{align}
and choose the conformal frame where the operators lie in the $(x^{+},x^{-})$ plane,
\begin{align}
x_1=(0,0), \qquad x_2=(z,\bar{z}), \qquad x_3=(1,1), \qquad x_4=\infty, \label{eq:conformal_frame}
\end{align}
where $(z,\bar{z})$ are the usual conformal cross-ratios (defined in \eqref{defzzbar}).
For the conformal blocks we also typically pull out a kinematic prefactor when working in a generic configuration to find a function of the cross-ratios:
\begin{align}
g^{\Delta_i}_{\Delta,\ell}(x_i)=\frac{1}{x_{12}^{\Delta_1+\Delta_2}x_{34}^{\Delta_3+\Delta_4}}\left(\frac{x_{24}}{x_{14}}\right)^{\Delta_1-\Delta_2}\left(\frac{x_{14}}{x_{13}}\right)^{\Delta_3-\Delta_4}g^{\Delta_i}_{\Delta,\ell}(z,\bar{z}).
\end{align}
The lightcone limit then corresponds to taking $$z\ll 1-\bar{z}\ll 1~.$$ To compute the spinning conformal block we can act with differential operators on the scalar block \cite{Costa:2011dw} and then take the lightcone limit. Although the blocks are not known in a simple closed form in general dimensions, they can be found in the lightcone limit using the Casimir equation \cite{DO3,Simmons-Duffin:2016wlq}. The construction of the spinning blocks is technical, but straightforward, so we will leave the details for Appendix \ref{app:spinningdetails}. The final result for the correlator in the lightcone limit due to massive spin-two exchange is 
\begin{align}\label{spin2result}
&g^{T\O\O T}_M(z\ll 1-\bar{z}\ll1)\approx   -(\lambda^{(2)}_{T\mathcal{O} M})^2 \,\frac{z^{\frac{1}{2}(\Delta_M-2-d-\Delta_{\mathcal{O}})}}{(1-\bar{z})^{\Delta_{\mathcal{O}}-d+2}}  
\nonumber \\ &\times\frac{\pi  (d-2)^2 \,(d-\Delta_{M}+\Delta_{\mathcal{O}})^2 (d-\Delta_{M}+\Delta_{\mathcal{O}}+2)^2\,\Gamma (\Delta_{M}+2)  \csc (\pi  (d-\Delta_{\mathcal{O}}))}{ \left((\Delta_{M}-\Delta_{\mathcal{O}})^2-d \left((\Delta_{M}-\Delta_{\mathcal{O}})^2-2\right)\right)^2 \Gamma (d-\Delta_{\mathcal{O}}-1) \Gamma^2 \left(\frac{1}{2} (-d+\Delta_{M}+\Delta_{\mathcal{O}}+4)\right)} \ .
\end{align}
This gives the left-hand side of \eqref{crosseq}. We remind the reader that we have dropped the ingredients necessary for extracting the $[\mathcal{O}\mathcal{O}]_{n,\ell}$ double-trace operators in the $t$-channel. As we see in a moment, once the kinematic prefactors are taken into account, this equation has exactly the right $\bar{z}$ dependence to contribute to the $[TT]_{0,\ell}$ double-traces in the $t$-channel.

To solve the bootstrap equation \eqref{crosseq}, we now analyze the right-hand side. To do so we need the conformal blocks $g^{\mathcal{O}\mathcal{O} TT}_{[TT]_{n,\ell}}(x_i)$. One way to do this is to act with the differential operators on a seed scalar block, make an Ansatz for the large spin OPE coefficients, and then perform the sum over spin at fixed twist. Here we will take a different route and instead decompose the $d$-dimensional spinning blocks in terms of the global $d=2$ conformal group, $SL(2,\mathbb{R})\times SL(2,\mathbb{R})$. The $d=2$ spinning blocks are simply a product of ${}_{2}F_1$ hypergeometrics \cite{Osborn:2012vt}. Physically, the $SL(2,\mathbb{R})$ group comes from transformations which preserves the light-ray separating two light-like operators. Since we are just keeping the leading dependence as $\bar{z}\rightarrow 1$, we can also work to leading order in the $d=2$ block expansion.  

 In $d=2$, we label the blocks by their holomorphic and anti-holomorphic weights $(h,\bar{h})$. These are related to the dimension and spin of the $d=2$ operators as 
 \begin{align}
 \Delta_{2d}=h+\bar{h}, \qquad \ell_{2d}=\bar{h}-h.
 \end{align}
To avoid confusion with the $d$-dimensional conformal dimension and spin we label all $d=2$ blocks by the weights $(h,\bar{h})$. Then, working in the same conformal frame as before, the $d=2$, $t$-channel expansion takes the form
\begin{align}
\<\mathcal{O}_1\mathcal{O}_{2}\mathcal{O}_3\mathcal{O}_4\>&=\frac{1}{(1-z)^{h_2+h_3}(1-\bar{z})^{\bar{h}_2+\bar{h}_3}}\sum\limits_{\chi_{h,\bar{h}}}\lambda_{23\chi_{h,\bar{h}}}\lambda_{14\chi_{h,\bar{h}}}g^{(d=2)}_{h,\bar{h}}(1-z,1-\bar{z}) ,
\\
g^{(d=2)}_{h,\bar{h}}(z,\bar{z})&=k_{h}^{h_{21},h_{34}}(z)k_{\bar{h}}^{\bar{h}_{21},\bar{h}_{34}}(\bar{z}) \ , ~~~~ k^{a,b}_{h}(z)\equiv z^{h}{}_{2}F_{1}(h+a,h+b,2h,z),
\end{align}
where $h_{ij}=h_i-h_j$. To specialize to our case we set $\mathcal{O}_1=\mathcal{O}_4=T^{--}$ and $\mathcal{O}_2=\mathcal{O}_3=\mathcal{O}$. Using \eqref{spin2result}, we now have to solve a bootstrap equation of the form
\begin{align}
\frac{z^{\frac{1}{2}(\Delta_{M}-2-d-\Delta_{\mathcal{O}})}}{(1-\bar{z})^{\Delta_{\mathcal{O}}-d+2}}\sim \frac{1}{((1-z)(1-\bar{z}))^{\Delta_{\mathcal{O}}}}\sum\limits_{\mathcal{O}}\lambda_{\mathcal{O}\mathcal{O}\chi_{h,\bar{h}}}\lambda_{T^{--}T^{--}\chi_{h,\bar{h}}}g^{(d=2)}_{h,\bar{h}}(1-z,1-\bar{z}) 
\end{align}
Comparing the $(1-\bar{z})$ dependence, we see the blocks on the right-hand side must have $\bar{h}\rightarrow d-2$ as $h\rightarrow \infty $.\footnote{As shown in \cite{Fitzpatrick:2012yx,Komargodski:2012ek} the sum over $h$ must be unbounded to match the power of $z$ on the left-hand side.} We can identify these as the minimal-twist, stress tensor double-trace operators, $T^{++}\partial^{+}...\partial^{+}T^{++}$. This is the claim advertised earlier: a $\< TT\O\>$ coupling in the dual CFT contributes, by way of solving crossing for $\< T\O\O T\>$, to the OPE coefficient $\<\mathcal{O}\mathcal{O}[TT]_{0,\ell}\>$.  

We may also read off the explicit form of the large-spin OPE coefficients:
\begin{align}
\label{eq:LargeSpinOPE}
&\lambda_{\mathcal{O}\mathcal{O}[T^{++}T^{++}]}\lambda_{T^{--}T^{--}[T^{++}T^{++}]} \sim (\lambda^{(2)}_{T\mathcal{O} M})^2 S_{\frac{1}{2}(\Delta_{M}-2-d-\Delta_{\mathcal{O}})}(h)  \\
&\times \frac{ (d-2)^2 (d-\Delta_{M}+\Delta_{\mathcal{O}}+2)^2(d-\Delta_{M}+\Delta_{\mathcal{O}})^2}{ \left((\Delta_{M}-\Delta_{\mathcal{O}})^2-d \left((\Delta_{M}-\Delta_{\mathcal{O}})^2-2\right)\right)^2 }
 \frac{ \Gamma (\Delta_{M}+2)  \Gamma(\Delta_{\O}-d+2)}{ \Gamma^2 \left(\frac{1}{2} (-d+\Delta_{M}+\Delta_{\mathcal{O}}+4)\right)} \ , \nonumber
\end{align}
where we have defined \cite{Simmons-Duffin:2016wlq},
\begin{align}
&S_{a}(h)=\frac{\Gamma^{2}(h)\Gamma(h-a-1)}{\Gamma^2(-a)\Gamma(2h-1)\Gamma(h+a+1)}.
\end{align}
As we are working at tree-level in $1/C_T$, the OPE coefficient $\lambda_{T^{--}T^{--}[T^{++}T^{++}]}$ is fixed to its MFT value.  Moreover, the three-point function $\<\mathcal{O}\mathcal{O} [TT]_{n,\ell}\>$ is fixed up to a single number and thus we do not lose any information by fixing the polarizations. The $\sim$ in (\ref{eq:LargeSpinOPE}) is a reminder this gives the asymptotic OPE coefficients in the limit $h\sim |\ell| \gg1$.\footnote{Since we are working with the chiral $d=2$ blocks, the spin can be negative.} To recover the case where we have bulk graviton exchange we can simply set $\Delta_{M}=d$.


\section{Local interactions} 
\label{sec:explicitcomputations2}
In this section, we will explicitly compute the corrections to the minimal-twist OPE coefficients $\<\mathcal{O}\mathcal{O}[TT]_{0,\ell}\>$ from interactions of the form (\ref{eq:ddphi2C2}). This section returns to the approach of \cite{Fitzpatrick:2019zqz}, by extracting these OPE coefficients from the heavy-light correlator $\< \O_H\O_H\O_L\O_L\>$, which we compute in the bulk. For concreteness we work in $d=4$. 

\subsection{Preliminaries}

We will compute $\< \O_H\O_H\O_L\O_L\>$ by solving the equations of motion for $\phi_L$ in order to obtain its bulk-to-boundary propagator, $\Phi$, in a fixed background created by the heavy state. For concreteness we refer to this background as a black hole, even though there exist backreacted geometries that are not thermal. The bulk light field $\phi_L$ is dual to the probe operator ${\cal O}_L$. Some basic relations are
\bea
\label{bEoM}
&&(-\nabla^2+ m^2) \Phi =0 \ , ~~ m^2= \Delta_L (\Delta_L -d)\ , \\
&&\Phi (r, x_1, x_2) =\langle {\cal O}_L (x_1) \phi_L (r,x_2) \rangle_{\rm{Black~hole}} \ ,\\
&&\lim_{r\to \infty} r^{\Delta_L} \Phi (r, x_1, x_2) = \langle  {\cal O}_H (0)  {\cal O}_L (x_1) {\cal O}_L (x_2)  {\cal O}_H (\infty) \rangle\ .
\eea
We denote $\Delta=\Delta_L$ and $\phi_L=\phi$ in the following. 

Solving the bulk field equation \eqref{bEoM} analytically is hard in general. Even in the Schwarzschild black hole background, no analytic solution is available. It is, however, possible to systematically obtain analytic solutions in a near-boundary expansion, which is all we need to study the conformal block decomposition of the boundary correlator.
In particular, it was shown in \cite{Fitzpatrick:2019zqz} that the lowest-twist, lowest-spin multi-$T$ operators 
\be\label{Tm}
[T^m] \equiv T_{\mu_1\nu_1}T_{\mu_2\nu_2}\ldots T_{\mu_m\nu_m},
\ee
i.e. the minimal-twist operators made from products of $T$ with no derivatives, are controlled by the leading large $r$ term $Q(w,u)$ in a solution for $\Phi$ of the form\footnote{This solution was considered for a planar black hole; see \cite{Fitzpatrick:2019efk} for the spherical black hole case.}
\be
\label{Q}
\Phi(r,w,u) = \Phi_{\rm{AdS}} \left( Q(w,u) + \CO(\frac{1}{r}) \right) \ ,
\ee 
where the variable $w$ can be defined via the pure AdS bulk-to-boundary propagator  
\be
\Phi_{\rm{AdS}}= \Big( {r\over w^2}\Big)^{\Delta} \ .
\ee  
In the planar black hole case with coordinates $(t,r, x_1, x_2, x_3)$,  $u= r^{-1}\sqrt{\sum_{i=1}^3  x^2_i}$. 
 In the large $r$ limit, \eqref{Q} is a solution to
\be
\nabla^2 \Phi = \Delta (\Delta-4) \Phi .
\ee
We can compute the corrections to the OPE coefficients in the same way, but including the additional interaction terms in the equation for $\phi$.

First, let us discuss why interactions of the form (\ref{eq:ddphi2C2}) are the simplest quartic terms that might lead to corrections to the minimal-twist multi-$T$ OPE coefficients.  The reason we need at least four derivatives is  that if we have only two derivatives, then their indices can only be contracted with $g_{\mu \nu}$ or $R_{\mu\nu}$.  But $R_{\mu\nu} \propto g_{\mu\nu}$ up to terms suppressed by $1/r^6$ in a large $r$ expansion (since this is the first order where we see that the metric is not AdS-Schwarzschild).  Moreover, the solution $\phi_0$ for $\phi$ in the absence of the new interaction terms satisfies
\be
\frac{\nabla^2 \phi_0}{\phi_0} \sim \CO(r^0), \qquad \frac{\nabla_\mu \nabla^\nu \phi_0}{\phi_0} \sim \CO(r^4)
\ee
at large $r$. Hence, the $\CO(1/r^6)$ and higher terms in $R_{\mu\nu}$ are negligible at infinite $r$:
\be
\lim_{r \rightarrow \infty}  R^{\mu\nu} \nabla_\mu \nabla_\nu \phi_0 \propto  \nabla^2 \phi_0 =  \Delta(\Delta-4) \phi_0,
\ee
and thus including such a term  is equivalent to shifting the mass of $\phi$, $\delta m^2  \propto \Delta(\Delta-4)$, 
as far as the minimal-twist multi-$T$ OPE coefficients are concerned.

Note that by this type of argument, one can also easily see that $\phi W^2$ evaluated on the fixed heavy-state background has no effect through its contribution to the equations of motion for $\phi$, since $W^2$ vanishes at infinite $r$.

Therefore, the first candidate term for a correction term quadratic in $\phi$ has four derivatives, and two Weyl tensors; a single Weyl tensor contributes to the equations of motion through $W^{\mu\nu \rho \sigma} \nabla_\mu \nabla_\nu \nabla_\rho \nabla_\sigma \phi$, which vanishes due to symmetry/anti-symmetry of the indices. A basis for such operators,  introduced in \cite{Policastro:2008hg}, is
\bea
\label{I1}
I_1^D &=& \nabla_m \nabla^n \phi \nabla_p \nabla^q \phi  W^{i m  j p} W_{injq}, \\ 
 I_2^D &=& \nabla_m \nabla_n \phi \nabla^p \nabla^q \phi  W^{i m  j n} W_{ipjq}, \\ 
 I_3^D &=& \nabla_m \nabla^n \phi \nabla_p \nabla^q \phi  W^{mpij} W_{ijnq}, \\ 
 I_4^D &=& \nabla_m \nabla_i \phi \nabla^i \nabla^n \phi  W^{mjkl} W_{njkl}, \\ 
\label{I5}
 I_5^D &=& \nabla_i \nabla_j \phi \nabla^i \nabla^j \phi  W^{klmn} W_{klmn}.
 \eea  
We have replaced the Riemann tensor in \cite{Policastro:2008hg} with the Weyl tensor in order to get rid of contributions that begin lower than $\CO(h^2)$ in an expansion in $h_{\mu\nu}$ (since lower orders can be absorbed into the mass and normalization of $\phi$ by using the equations of motion). The authors of \cite{Policastro:2008hg} found that in type IIB supergravity, only $I_1^D$ is generated, and its coefficient in the Lagrangian is 
\be
\frac{\alpha'^3}{2 \kappa^2} \zeta(3),
\ee
where $\kappa^2 = 8 \pi G_N$. 

To get some sense of which of the $I_n^D$ we might expect to affect the minimal-twist OPE coefficients, we may look at their contribution to the $S$-matrix for $\phi h \rightarrow \phi h$ scattering.  
Since the $[TT]_{0,\ell}$ trajectory starts at $\ell=4$, we only need to consider amplitudes contributing as $\ell\geq 4$ in the $s$-channel, $ \phi \phi \rightarrow h h$. Computing the tree-level 2-to-2 amplitudes for the $I_n^D$ interactions is straightforward.  The linear approximation of the Riemann tensor is
\be
R_{\mu\nu\rho\sigma} \rightarrow -\frac{1}{2} \left( p_\rho p_\nu \epsilon_{\mu\sigma} + p_\sigma p_\mu \epsilon_{\nu \rho} - p_\sigma p_\nu \epsilon_{\mu \rho} - p_\rho p_\mu \epsilon_{\nu \sigma} \right) ,
\ee
where $p$ is the graviton momentum and $\epsilon_{\mu\nu}$ is its polarization.   
 Now label the two $\phi$ momenta $p_1$ and $p_2$, and the two graviton momenta and polarizations $p_3, p_4$ and $\epsilon_3, \epsilon_4$, respectively.  The ingredients for the $S$-matrix are
\begin{align}
s=-(p_1+p_2)^{2}, \quad t=-(p_1+p_3)^{2}, \quad \epsilon_3\cdot\epsilon_4, \quad
\epsilon_i\cdot p_j
\end{align}
Here we are writing the graviton polarization tensors as $\epsilon^{\mu\nu}_{i}=\epsilon^{\mu}_i\epsilon^{\nu}_i$ where $\epsilon$ is null and we can set $\epsilon_i\cdot p_i=0$.  In terms of Mandelstam invariants, only terms containing $(\epsilon_3 \cdot \epsilon_4)^2 t^{4}$ can affect spin-4 operators, so lower powers of $t$ can be discarded for our purposes.\footnote{As shown in  \cite{Camanho:2014apa}, terms with other dot products, like $\epsilon_i\cdot p_j$ are subleading in the Regge limit and can be ignored (see the discussion around (B.2) there). }  Substituting the linearized approximation of the Riemann tensor into the $I_n^D$ interactions and keeping only $(\epsilon_3 \cdot \epsilon_4)^2 t^4$ terms, it is easy to obtain
\bea
I_1^D \cong \frac{1}{2^6} t^4 (\epsilon_3 \cdot \epsilon_4)^2\ , ~~~ 
I_2^D  \cong \frac{1}{2^6} t^4 (\epsilon_3 \cdot \epsilon_4)^2 \ , ~~~
I_3^D \cong  0\ , ~~~
I_4^D \cong  0\ , ~~~
I_5^D \cong  0  \ .
 \label{eq:IDSmat}
\eea

Based on (\ref{eq:IDSmat}), we expect that $I_i^D$ should affect the $[TT]_{0,\ell}$ OPE coefficients for $i=1,2$ but not for $i=3,4,5$.  


 \subsection{Explicit OPE coefficients of $T^2$}
\label{s42}
 
Having set our expectations for what we should find, we next present the results and some of the details of the actual computation for the calculation of OPE coefficients of $T^2$ operators coming from the addition of the local operators $I_n^D$. We add to the Lagrangian 
\be
\delta \CL ={1\over 2} \sum_i a_i I_i^D,
\ee
and  consider their contribution to the equations of motion:
\be
\left( -\nabla^2 +m^2  \right) \phi + \frac{\partial \delta {\cal S}}{\partial \phi} = 0,
\ee
where $m^2 = \Delta(\Delta-4)$.  The background metric created by the heavy state is 
 \be \label{eq:BHMetric}
ds^2 =  
\big(1+r^2 f(r)\big) dt^2 + \frac{dr^2}{1+r^2 h(r)} + r^2 d\Omega_{3}^2 \ ,
\ee
with
\be\label{fandh}
f(r) =1 - \frac{f_0}{r^4} - \frac{f_4}{r^8} + \dots \ , ~~~  h(r) = 1- \frac{h_0}{r^4} - \frac{h_4}{r^8} + \dots,   
 \ee
 where  $\dots$ indicate higher-order terms in $1/r$ and we have set the AdS radius to unity. Conformal invariance requires $h_0=f_0$ \cite{Fitzpatrick:2019zqz}, while the $h_4$ and $f_4$ terms parameterize further departures from global AdS.\footnote{As these $I_i^D$ interactions do not affect the line element at large $C_T$, they are unseen in the geodesic.  
Note, however, the coeffcients of higher-curvature terms $f_4, h_4$ etc can affect the geodesic; see \cite{Fitzpatrick:2019zqz} for detailed expressions.}  In terms of the quantum numbers $\D_H$ and $C_T$, 
\be
 f_0= {160\over 3} {\Delta_H\over C_T} \ .
\ee 

Here we consider a spherical black hole and will adopt the planar limit for a simpler all-order analysis later. We take an Ansatz for the bulk-to-boundary propagator $\Phi$ of the form
\be
\Phi = \Phi_{\rm AdS} \times \left(1 + \sum_{n=1}^\infty \frac{g_{2n}(r \sinh t, r \sin \theta)}{r^{2+2n}(1+r^2 (\sinh^2 t + \sin^2 \theta))^n} \right)
\ee
where $g_{2n}$ is an even, degree $(2n+2)$ polynomial in its arguments.\footnote{The most efficient way we have found  to solve for the  $g_{2n}$s is to use  Ansatzes that are meromorphic functions of $\Delta$ of the following form:
\be
g_{2n}(x,y) = \sum_{j=1}^{n+1} \frac{r_{n,j}(x,y)}{\Delta-j} + \sum_{i=0}^{\lfloor \frac{n-1}{2} \rfloor} p_{n,i}(x,y) \Delta^i
\ee  For more details, see \cite{Fitzpatrick:2019zqz}.   
} 
We have used the rotational symmetry to remove dependence on two angular coordinates. The equations of motion can be solved order-by-order in a large $r$ expansion, with $r \sinh t$ and $ r \sin \theta$ held fixed. 
Note that the $[T^m]$ conformal blocks are insensitive to the horizon boundary condition as observed in \cite{Fitzpatrick:2019zqz}.  
Moreover, we restrict to non-integer $\Delta$ to avoid mixing with double-trace operators made from two probe operators;  the mixing with double-trace modes will be indicated by poles at integer $\Delta$ in the OPE coefficients below.

The new terms $\delta \CL$ do not affect the equations for $g_2$ and $g_4$, and instead contribute first at $g_6$.  Once we have the bulk-to-boundary propagator, we obtain the boundary heavy-light correlator by taking $r\rightarrow \infty$ with $t, \theta$ fixed, 
\be
\< \CO_H \CO_H \CO_L \CO_L\> = \lim_{r\rightarrow \infty} r^{-\Delta} \Phi(t,\theta,r).
\ee
To obtain the OPE coefficients, we perform a conformal block decomposition. To use the standard form of the conformal blocks, we change to $z, \bar{z}$ coordinates, 
  \be
z = e^{t+ i \theta}, \quad \quad  \bar{z} = e^{t-i \theta}, 
\label{eq:tthetaTozzbar2}
\ee
 and decompose the correlator as
\be
\<  \CO_H \CO_H \CO_L \CO_L\>  = \sum_{\Delta, \ell} p_{\Delta, \ell} g_{\Delta, \ell}(1-z,1-\bar{z}),
\ee 
where 
\be
p_{\Delta,\ell}\equiv \lambda_{\O_H\O_H \O_{\D,\ell}}\lambda_{\O_L\O_L \O_{\D,\ell}}
\ee
The single-stress tensor exchange contribution is
\be
\label{p42}
p_{4, 2}= f_0 {\Delta\over 120 } \ .
\ee

We are interested in the OPE coefficient $\lambda_{\O_L\O_L \O_{8,4}}$ for the minimal-twist composite operator $\O_{8,4} \equiv [TT]_{0,4} = T_{\mu\nu}T_{\rho\sigma}$, appearing in the product
\be
p_{8,4} = \lambda_{\O_H\O_H \O_{8,4}}\lambda_{\O_L\O_L \O_{8,4}} \ .
\ee
Following this procedure, we have found that $p_{8,4}$ is corrected by an $(a_1+a_2)$ piece:
 \bea
\label{p84}
p_{8,4} &=&{f_0^2\over \D-2} \Big( \frac{ \Delta (7 \Delta ^2+6 \Delta +4) }{201600}  -\left(a_1+a_2\right)  \frac{   \Delta (\Delta +1) (\Delta +2) (\Delta +3)}{462}\Big).
 \eea 
As we expect based on (\ref{eq:IDSmat}), only the terms $I_i^D$ for $i=1,2$ correct $p_{8,4}$.  The desired OPE coefficient $\lambda_{\O_L\O_L \O_{8,4}}$ may be read off from the above and contains the interesting functional dependence, since the heavy ``half'' of $p_{8,4}$ simply contributes the $f_0^2$ times an overall constant -- that is, $\lambda_{\O_H\O_H \O_{8,4}}$ is (obviously) $\D$-independent, and is proportional to $\D_H^2/C_T = f_0^2 C_T$.\foot{One can see this via Witten diagrams as in section \ref{sec:WittenDiagramsExchange}: extracting this from $\< T \O_H\O_H T\>$, the exchange $T\O_H \rar \O_H \rar T\O_H$ is proportional to $\D_H^2$ by the Ward identity \eqref{ward} and survives at $\D_H\gg 1$.} We observe that the correction term $\sim (a_1+a_2)$ dominates at large $\Delta$; and that the $I^i_D$ corrections vanish at $\Delta=-1,-2, -3$. 

As an aside, note that the OPE coefficients for  higher-twist multi-traces become more sensitive to model-dependent parameters.\footnote{
The sub-leading- and sub-sub-leading-corrections, $\delta p_{8,2}$ and $\delta p_{8,0}$, due to $I^i_D$ are given by 
{\scriptsize
    \bea
  && \delta p_{8,2} = -f_0^2  \Delta  (\Delta +1) (\Delta +2)  \frac{\left(13 a_1+13 a_2+44 a_4\right) (\Delta ^2-4\Delta)-53 a_1-141 a_2+1672 a_3-792 a_4}{6930 (\Delta -3) (\Delta -2)} , \\
    && \delta p_{8,0} = -\frac{f_0^2  \Delta  (\Delta +1)  }{13860 (\Delta -4) (\Delta -3) (\Delta -2)} \left[3 \left(23 a_1+23 a_2+110 a_4+594 a_5\right) (\Delta ^4-8\Delta^3) \right. \nn\\
&&\left. ~~~~~~~~~~ +\left(1753 a_1-2581 a_2+4686 a_3-6952 a_4-35640
   a_5\right) \Delta ^2  -44 \left(167 a_1-443 a_2+426 a_3-1220 a_4-5832 a_5\right) \Delta\right. \nn\\
   &&\left.~~~~~~~~~~ -24 \left(491 a_1-1973 a_2+1848 a_3-7601 a_4-23760 a_5\right)\right]  \ .
\eea}} 
The full $p_{8,0}$ and $p_{8,2}$ can also depend on the coefficients of higher-derivative terms ($\sim f_4, h_4$) in the bulk action; see \cite{Fitzpatrick:2019zqz} for explicit expressions.

Let us also comment on the case of 4d $\mathcal{N}=4$ super-Yang-Mills (SYM). The fact that the minimal-twist OPE data is corrected by the structure $I_1^D$ implies that for $\mathcal{N}=4$ SYM, for which the AdS$_5$ effective action contains such a term \cite{Policastro:2008hg}, the correction from the terms considered herein is nonzero. This is perhaps suggestive that even in very special CFTs, minimal-twist OPE data receives perturbative corrections in $1/\D_\gap$, i.e. $\ell_s \sim \l^{-1/4}$ in the $\mathcal{N}=4$ SYM case.


 \subsection{An all-order analysis via a planar black hole}

In this subsection, we now study the effect of the interactions $I_n^D$ given in \eqref{I1}-\eqref{I5} on the minimal-twist OPE coefficients for the $[T^m]$ operators defined in \eqref{Tm}. By developing an algorithm for computing these, we prove that only $I_1^D$ and $I_2^D$ contribute, and that they never generate $f_{i>0}$-or $h_{i>0}$-dependence. We give several lowest-twist OPE coefficients explicitly and observe certain patterns. To keep the analysis simple, we here focus on the planar (i.e. high-temperature) black hole limit, which is sufficient to extract lowest-twist data, and identify patterns.

Consider a planar black hole with the metric
\be
ds^2= r^2 f(r) dt^2  + {dr^2 \over r^2 h(r) } + r^2 \sum_{i=1}^{3}dx_i^2\ ,
\ee 
where black-hole solutions $f(r)$ and $h(r)$ depend on the details of a theory. 
Near the boundary, $f(r)$ and $h(r)$ are parameterized as in \eqref{fandh}.  
 The additional bulk couplings \eqref{I1}-\eqref{I5} lead to a modified equation of motion:
\bea
&&\big(\nabla^2 - m^2 -  a_1   \nabla^n  \nabla_m \nabla_p \nabla^q \phi  W^{i m  j p} W_{injq} \nn\\
&&~~~~~~~~~~~~~ -  a_2    \nabla_n \nabla_m \nabla^p \nabla^q \phi  W^{i m  j n} W_{ipjq}\nn\\
&&~~~~~~~~~~~~~ -  a_3    \nabla^n \nabla_m \nabla_p \nabla^q \phi  W^{mpij} W_{ijnq}\nn\\
&&~~~~~~~~~~~~~ -  a_4     \nabla_i  \nabla_m \nabla^i \nabla^n \phi  W^{mjkl} W_{njkl}\nn\\
&&~~~~~~~~~~~~~ -  a_5   \nabla_j   \nabla_i \nabla^i \nabla^j \phi  W^{klmn} W_{klmn}\big) \phi =0 \ .
\eea
After performing integration by parts in varying the action, we have dropped covariant derivatives of Weyl curvatures as they are subleading in a large $r$ expansion. 
 Following \cite{Fitzpatrick:2019zqz}, we adopt the form \eqref{Q}. With fixed $w,u$, the lowest-twist sector can be isolated in a large $r$ limit.  
We find the following {\it reduced} bulk field equation that captures the lowest-twist sector of the dual CFT: 
\bea
\label{newrEoM}
&& u^{-2} (1-w^2)^{-1} \partial_w \Big( w^{1-2\Delta} (1-w^2)^{2} \partial_w Q \Big) +  u^{-1} k_{-}  \partial_w \Big( w^{-2\Delta} k_{+} \partial_u Q\Big)  \\
&& - w^{1-2\Delta} u \partial_u \Big(u^{-1} \partial_u Q \Big)+ f_0 \partial_w \Big( w^{-1} \partial_w (w^{-2\Delta} Q)\Big) = - 24 \big(a_1+ a_2\big) f^2_0  u^2 w^{-7-2\Delta} ~ \sum_{j=0}^{4} \alpha_j \partial_w^{(j)}  Q \nn
\eea 
where
\bea
&& k_{\pm}= \big(2 w^2-4 \big)^{{1\over 2} \pm {\Delta\over 2}} \ , \\
\label{alpha0}
&& \alpha_0=  16 \Delta \big(\Delta+1\big) \big(\Delta+2\big)\big(\Delta+3\big) \ , \\
&& \alpha_1= -w \big(5 + 4 \Delta\big) \Big(3 + 4 \Delta (5 + 2 \Delta)\Big)\ , \\
&& \alpha_2 =  3 w^2 \Big(5 + 8 \Delta (2 + \Delta)\Big) \ , \\
&& \alpha_3 =  -2 w^3 \big(3 + 4 \Delta\big) \ , \\
&& \alpha_4 =  w^4  \ .
\eea 
The result \eqref{newrEoM} implies that, to {\it all} orders in the power of $f_0$, neither $a_3, a_4, a_5$ nor higher-curvature corrections $f_{i>0}, h_{i>0}$ enter into the lowest-twist sector of the $T^m$ operators. 

We can solve the lowest-twist bulk field equation \eqref{newrEoM} order-by-order in twist. 
The solution has the following structure:
\be
Q(w,u)=1+  \sum_{k=-2}^2 q_{k,2} w^k u^2+  \sum_{k=-6}^4 q_{k,4} w^k u^4+  \sum_{k=-8}^6 q_{k,6} w^k u^6 + \cdots
\ee 
where $k$ increases by 2. The relation between $q_{k,k}$ and the lowest-twist OPE coefficients $p_{\Delta ,J}$ reads 
\be
q_{k,k}=(-4)^{k\over 2} p_{2k, k}.
\ee
where we remind the reader that $p_{2k,k}$ is the product of OPE coefficients 
\be
p_{2k,k} \equiv \lambda_{\O_H\O_H [T^{k\o 2}]}\l_{\O_L\O_L [T^{k\o 2}]}\,,
\ee
where $[T^{k\o 2}]$ were defined in \eqref{Tm}. We obtain these by computing coefficients $q_{k,2}, q_{k,4}, q_{k,6},...$, and then extracting the lowest-twist members $p_{2k,k}$ by taking the boundary limit.

The results $p_{4, 2}$ and $p_{8, 4}$ were given in \eqref{p42} and \eqref{p84}, respectively. 
The  equation \eqref{newrEoM} allows one to compute the lowest-twist OPE coefficients efficiently.
At the next order, 
\bea
\label{p126}
&&p_{12, 6}= {\Delta  f_0^3\o (\D-2)(\D-3)} \Big(\frac{  1001 \Delta^4 +3575 \Delta^3 +7310\Delta^2 +7500 \Delta +3024}{10378368000}\nn\\
&&~~~~~ - (a_1+a_2)  (\Delta+1) (\Delta+2) (\Delta+3)   { 84 + \Delta (53 + 13 \Delta) \over 
 720720 } \Big)\ . 
\eea 
It is straightforward to compute higher-order coefficients.\footnote{For instance, 
{\scriptsize
\bea
&&p_{16,8}=f_0^4 \Delta \frac{\Delta  (\Delta  (\Delta  (17 \Delta  (1001 \Delta  (7 \Delta +57)+246150)+10867340)+16958856)+14428176)+5009760}{592812380160000 (\Delta -4) (\Delta -3) (\Delta -2)}\nn\\
&&+(a_1+a_2)^2  f_0^4\Delta  (\Delta +1) (\Delta +2) (\Delta +3) \frac{\Delta  (\Delta  (\Delta  (\Delta  (4199 \Delta +80683)+698029)+3253937)+7918932)+7893900}{1792502712 (\Delta -4) (\Delta -3) (\Delta -2)}\nn\\
&&-(a_1+a_2)  f_0^4\Delta  (\Delta +1) (\Delta +2) (\Delta +3) \frac{\Delta  (\Delta  (323 \Delta  (91 \Delta +911)+1368692)+3215632)+3145800 }{391091500800 (\Delta -4) (\Delta -3) (\Delta -2)}\ .
\eea}} As explained below \eqref{p84}, here and for general $k$, the interesting dependence comes from the $\l_{\O_L\O_L [T^{k\o 2}]}$ half of the product of OPE coefficients, while $\l_{\O_H\O_H [T^{k\o 2}]}$ simply contributes the factor $f_0^{k\o 2}$, which is visible in \eqref{p126}.   

Let us close with a few observations. Again, the correction term  dominates at large $\Delta$, and the corrections vanish at $\Delta=-1,-2, -3$. We have explicitly checked that this form persists to much higher orders. We do not have a complete intepretation for these special values of $\Delta$ that enhance the universality, but a possibility is that they may correspond to certain null states in this class of theories.\foot{Note also that $\alpha_0$ does not dictate the lowest-twist  OPE coefficients in the boundary limit although the bulk coefficient  $\alpha_0$ in \eqref{alpha0} vanishes at $\Delta=-1, -2, -3$.  Namely, the universal factor $\big(\Delta+1\big) \big(\Delta+2\big)\big(\Delta+3\big)$ would still appear in the lowest-twist OPE coefficients if one sets $\alpha_0=0$.}


\section{Discussion}
\label{sec:discussion}

In this work, we have generalized the previous analysis \cite{Fitzpatrick:2019zqz} of stress-tensor composite dynamics in $d>2$ conformal theories at large $C_T$.  
By incorporating a larger set of possible terms in the OPE, we have shown that the minimal-twist multi-stress tensor sector can depend on the gap scale of AdS effective field theory. 
This result likewise implies that the heavy-light correlators in general can depend on the gap scale.\foot{Our heavy-light discussion focuses on $d=4$ but we expect similar corrections in other dimensions $d>2$.} In the process, we explained why the minimal-twist OPE coefficients are independent of the purely gravitational action. 

It may be interesting to develop CFT methods to systematically include the additional data the minimal-twist OPE coefficients depend on. Relatedly, it would be interesting to understand how to extend the near-lightcone Ansatz of \cite{Karlsson:2019dbd, Karlsson:2020ghx, Parnachev:2020fna} for heavy-light correlators in $d>2$; the results here suggest that the Ansatz must be generalized when moving away from CFTs at strictly infinite gap. It may be also potentially useful to ask, for certain special CFTs, whether there exists an algebraic structure that governs near-lightcone dynamics in $d>2$, and if it is possible to incorporate additional parameters for more general cases (see \cite{Huang:2020ycs} for a recent discussion).

Obtaining closed forms of minimal-twist OPE coefficients at finite spin due to a $\phi W^2$ bulk term could be useful. A method for doing this using spinning conformal blocks was described in section \ref{s3}. In the bulk approach, one has to solve for a new black-hole solution, taking the backreaction of $\phi$ into account. Both methods present technical challenges but it could be interesting to study these OPE coefficients in more detail. We would also like to understand more precisely the behavior of OPE coefficients at certain negative integer values of the scaling dimension observed in section \ref{sec:explicitcomputations2} and find implications of these zeros. 

There are other extensions to consider: for example, higher-order expansion in the lightcone limit, or the couplings to minimal-twist $[T^m]$ composites with $m>2$ and spin $\ell>2m$. It may be also worth extending the computation to include a shockwave and find corrections to the OPE coefficients in the Regge limit; the analysis in the absence of bulk matter fields was considered recently in \cite{Fitzpatrick:2019efk}. Finally, the discussion in section \ref{s42} and \cite{Policastro:2008hg} imply that one should find a gap-scale dependence in the lightcone correlators of $d=4$, $\mathcal{N}=4$ SYM. A more careful look at this question is warranted.

\newpage

\begin{center}
\subsection*{Acknowledgments}
\end{center}

ALF and KWH were supported in part by the US Department of Energy
Office of Science under Award Number DE-SC0015845 and in part by the Simons Collaboration Grant on the Non-Perturbative Bootstrap, and ALF in part by a Sloan Foundation
fellowship. The research of DM was supported by the Walter Burke Institute for Theoretical Physics and the Sherman Fairchild Foundation. 
EP was supported by Simons Foundation grant 488657.  DSD was supported in part by Simons Foundation grant 488657 (Simons Collaboration on the Nonperturbative Bootstrap), a Sloan Research Fellowship, and a DOE Early Career Award under grant no. DE-SC0019085. This work was initiated at the Aspen Center for Physics, which is supported by National Science Foundation grant PHY-1607611. 
\vskip .2 in

\appendix
\section{Spinning blocks}
\label{app:spinningdetails}
In this appendix we fill in the details to find the exchange $T\phi \rightarrow M\rightarrow \phi T$ in the lightcone limit. We will follow the procedure given in \cite{Costa:2011mg,Costa:2011dw}. To start, we recall the three-point function takes the form
\begin{align}
\<T \mathcal{O} \chi_{\Delta,\ell}\>=\frac{1}{P_{12}^{\Delta_{T\mathcal{O} \chi}}P_{23}^{\Delta_{\mathcal{O}  \chi T}}P_{13}^{\Delta_{T \chi \mathcal{O} }}}\sum\limits_{i=0}^{2}\lambda^{(i)}_{T\mathcal{O} \chi}k^{2-i}_{1}k_{3}^{\ell-i}m_{13}^i.
\end{align}
To construct the spinning block, we want to rewrite this as a differential operator acting on a seed scalar structure, $\<\mathcal{O}_1\mathcal{O}_2\chi_{\Delta,\ell}\>$. The basic differential operators we need are\footnote{In $D_{11}$ we dropped terms which vanish when acting on $\<T \mathcal{O}  \chi_{\Delta,\ell}\>$. }
\begin{align}
D_{11}&=(P_1\cdot P_2)(Z_1\cdot \frac{\partial}{\partial P_2})-(Z_1\cdot P_2 )(P_1\cdot\frac{\partial}{\partial P_2})\ ,
\\
D_{12}&=(P_1\cdot P_2)(Z_1\cdot \frac{\partial}{\partial P_1})-(Z_1\cdot P_2)(P_1\cdot \frac{\partial}{\partial P_1})+(Z_1\cdot P_2)(Z_1\cdot \frac{\partial}{\partial Z_1}) \ .
\end{align}
If we define the dimension shifting operator $\Sigma^{ij}$ which acts as $(\Delta_1,\Delta_2)\rightarrow (\Delta_1+i,\Delta_2+j)$ then the differential basis is given by
\begin{align}
\<T \mathcal{O} \chi_{\Delta,\ell}\>=\sum\limits d^{(i)}_{T\mathcal{O}  \chi}D_{11}^iD_{12}^{2-i}\Sigma^{i,2-i}\frac{1}{P_{12}^{\Delta_{\mathcal{O}_T \mathcal{O}  \chi}}P_{23}^{\Delta_{\mathcal{O} \chi \mathcal{O}_T}}P_{13}^{\Delta_{\mathcal{O}_T \chi\mathcal{O} }}}.
\end{align}
Here $\mathcal{O}_T$ is a fictitious scalar operator with dimension $\Delta_{\mathcal{O}_T}=d$. The change of basis between the standard and differential basis is
\begin{align}
d^{(i)}_{T\mathcal{O} M}&=\sum\limits_{j=0}^{2}R^{i}_{j}\lambda^{(j)}_{T\mathcal{O} M},
\\
R&=\left(
\begin{array}{ccc}
 \frac{1}{(\Delta_{M}-1) \Delta_{M}} & \frac{-d+\Delta_{\mathcal{O}}+\Delta_{M}+2}{2 \Delta_{M}(1- \Delta_{M})} & \frac{(-d+\Delta_{\mathcal{O}}+\Delta_{M}) (-d+\Delta_{\mathcal{O}}+\Delta_{M}+2)}{2 (\Delta_{M}-1) \Delta_{M}} \\
 \frac{2}{(\Delta_{M}-1) \Delta_{M}} & \frac{-d+\Delta_{\mathcal{O}}+2}{\Delta_{M}(1-\Delta_{M})} & \frac{(-d+\Delta_{\mathcal{O}}-\Delta_{M}+2) (-d+\Delta_{\mathcal{O}}+\Delta_{M})}{(\Delta_{M}-1) \Delta_{M}} \\
 \frac{1}{(\Delta_{M}-1) \Delta_{M}} & \frac{d-\Delta_{\mathcal{O}}+\Delta_{M}-2}{2 (\Delta_{M}-1) \Delta_{M}} & \frac{(d-\Delta_{\mathcal{O}}+\Delta_{M}-2) (d-\Delta_{\mathcal{O}}+\Delta_{M})}{2 (\Delta_{M}-1) \Delta_{M}} \\
\end{array}
\right). \label{eq:changeofbasis}
\end{align}
To keep the notation compact, we define
\begin{align}
\mathcal{D}_{L}=\sum\limits d^{(i)}_{T\mathcal{O} \chi}D_{11}^iD_{12}^{2-i}
\end{align}
where we use the change of basis (\ref{eq:changeofbasis}) and the conservation constraints (\ref{eq:ConsTphiV}). The subscript $L$ is to remind us this acts on the points $(x_1,x_2)$. We can define $\mathcal{D}^{T\mathcal{O}}_{R,V}$ by letting $1\rightarrow 4$ and $2\rightarrow 3$. These operators naturally act on the conformal block as a function of all four positions, but as before it is simplest to pull out an overall kinematic prefactor
\begin{align}
g^{\Delta_i}_{\Delta,\ell}(x_i)=\frac{1}{x_{12}^{\Delta_1+\Delta_2}x_{34}^{\Delta_3+\Delta_4}}\left(\frac{x_{24}}{x_{14}}\right)^{\Delta_1-\Delta_2}\left(\frac{x_{14}}{x_{13}}\right)^{\Delta_3-\Delta_4}g^{\Delta_i}_{\Delta,\ell}(z,\bar{z}) \label{eq:prefactorblock}
\end{align}
where $(z,\bar{z})$ are defined as usual by
\begin{align} \label{defzzbar}
z\bar{z}=\frac{x_{12}^{2}x_{34}^{2}}{x_{13}^2x_{24}^2}, \qquad (1-z)(1-\bar{z})=\frac{x_{14}^{2}x_{23}^{2}}{x_{13}^2x_{24}^2}.
\end{align}
Then the spinning conformal block for the exchange $T \mathcal{O} \rightarrow M \rightarrow \mathcal{O} T$ is
\begin{align}
g^{T\mathcal{O}\mathcal{O}T}_{M}(x_i)=\mathcal{D}_{L}\mathcal{D}_{R} g^{\mathcal{O}_T\mathcal{O}\mathcal{O} \mathcal{O}_T}_{M}(x_i).
\end{align}
To find the spinning block in the lightcone limit, it is sufficient to know the seed scalar block in the same limit. Taking the limit $z\ll1$ we have
\begin{align}
g^{\mathcal{O}_1...\mathcal{O}_4}_{\Delta,\ell}(z,\bar{z})\approx z^{\frac{\Delta-\ell}{2}}k^{a,b}_{\frac{\Delta+J}{2}}(\bar{z}), \qquad a=\frac{\Delta_{21}}{2}, \qquad b=\frac{\Delta_{34}}{2}. \label{eq:scalarblockLC}
\end{align}
We can then expand the $SL(2,\mathbb{R})$ block around $\bar{z}=1$ and we see powers of $(1-\bar{z})^{n}$ and $(1-\bar{z})^{-a-b+n}$ with $n$ integer. After taking into account the prefactors in (\ref{eq:prefactorblock}) these are matched in the $t$-channel by the double-twist operators $[\mathcal{O}_2\mathcal{O}_3]_{n,\ell}$ and $[\mathcal{O}_1\mathcal{O}_4]_{n,\ell}$, respectively. It is then straightforward to act with the differential operators on (\ref{eq:scalarblockLC}) to find the spinning block in the lightcone limit. Although the computations are carried out in embedding space, we find the final answer is simplest when we go to the conformal frame \eqref{eq:conformal_frame} and use null, $d$-dimensional polarization vectors $\epsilon_i$. Then in the lightcone limit $z\ll 1-\bar{z}\ll1$ the spinning block is given by
\begin{align}
g^{T\mathcal{O}\mathcal{O} T}_{M}(x_i)\approx (\lambda^{(2)}_{T\mathcal{O} M})^2& \frac{z^{\frac{1}{2}(\Delta_M-2-d-\Delta_{\mathcal{O}})}}{(1-\bar{z})^{\Delta_{\mathcal{O}}-d+2}} (\epsilon_1^+\epsilon_2^+)^2 \frac{  (d-2)^2 \Gamma (\Delta_{M}+2)}{ \left((\Delta_{M}-\Delta_{\mathcal{O}})^2-d \left((\Delta_{M}-\Delta_{\mathcal{O}})^2-2\right)\right)^2 }
\nonumber \\ &\frac{ (d-\Delta_{M}+\Delta_{\mathcal{O}})^2 (d-\Delta_{M}+\Delta_{\mathcal{O}}+2)^2\Gamma(\Delta_{\O}-d+2) }{\Gamma^2 \left(\frac{1}{2} (-d+\Delta_{M}+\Delta_{\mathcal{O}}+4)\right)}+... 
\end{align}
where as a reminder this is the full conformal block without the kinematic prefactor in (\ref{eq:prefactorblock}) pulled out and we have dropped terms which are mapped to $[\mathcal{O}\mathcal{O}]_{n,\ell}$ in the $t$-channel. More precisely, this is the leading lightcone contribution of $M$ to the correlator $\<T^{--}\O\O T^{--}\>$.


\section{Comments on Loops}
\label{app:Loops}

We can ask how the results we found in this work, where we see corrections to minimal-twist universality via the lightcone bootstrap, are consistent with previous work \cite{Li:2019zba, Karlsson:2020ghx}, where universality was sought via the inversion formula. The two methods must agree when we study OPE data for operators of asymptotically large spin. In this appendix, we will show how the model-dependence of minimal twist OPE data can also be found by studying the inversion formula for $\<\O\O\O\O\>$.

\begin{figure*}[h]
    \centering
    \begin{subfigure}[t]{0.4\textwidth}
        \centering
        \includegraphics[scale=.27]{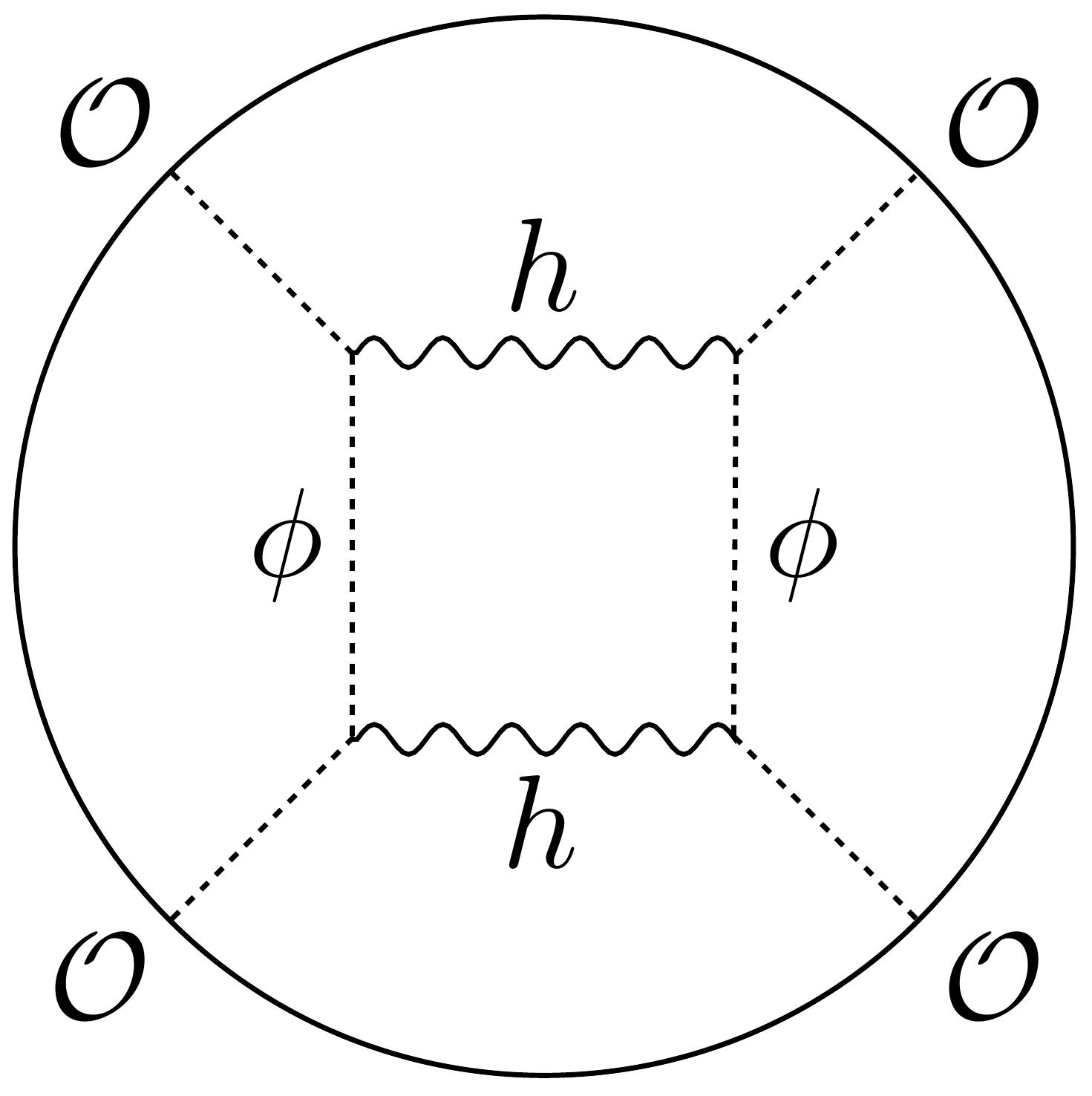}
        \caption{}
        \label{fig:universal_graviton_loop}
    \end{subfigure}
    ~ 
    \begin{subfigure}[t]{0.4\textwidth}
        \centering
        \includegraphics[scale=.27]{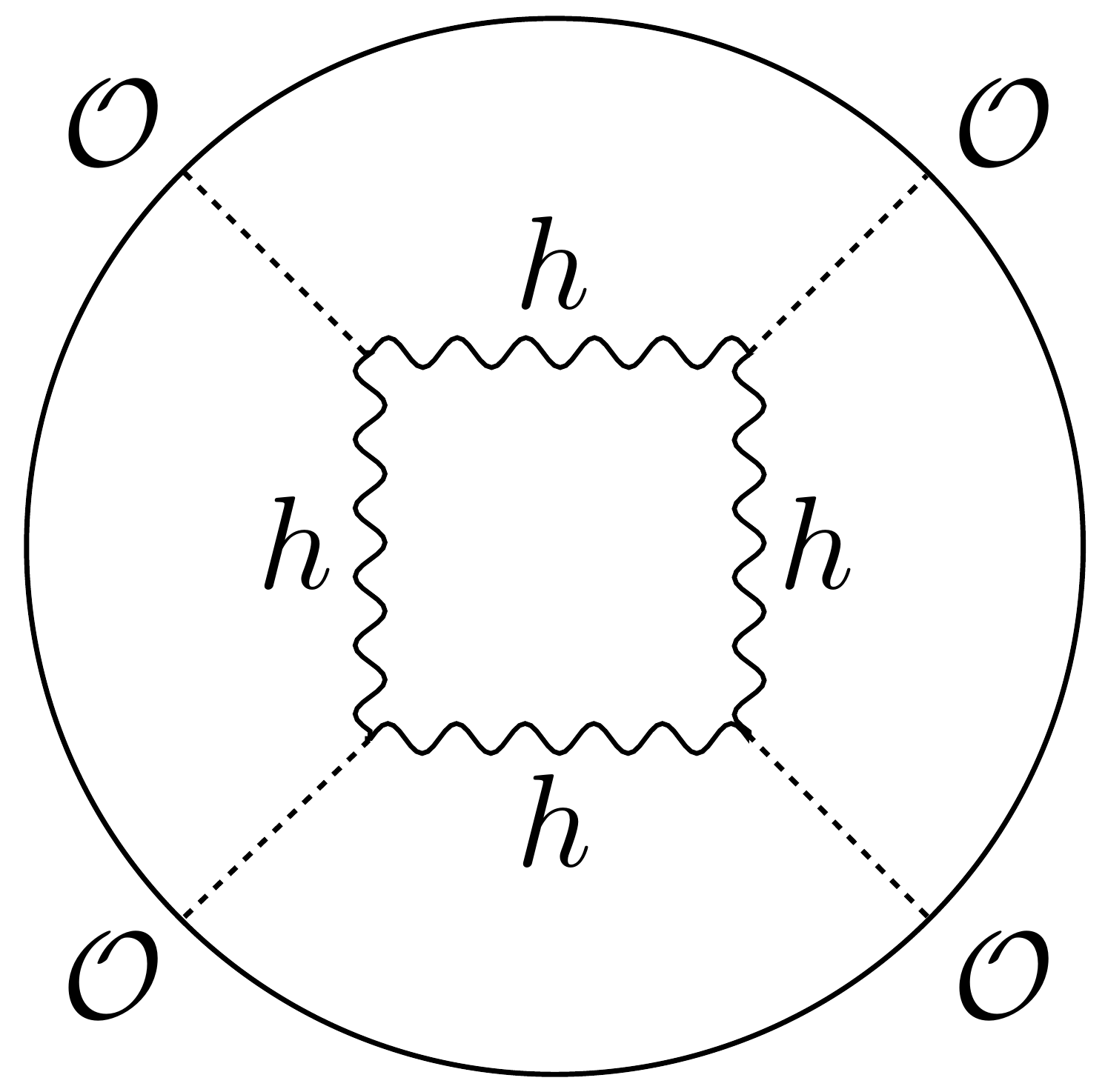}
        \caption{}
        \label{fig:purely_graviton_loop}
    \end{subfigure}
    \caption{}
\end{figure*}

To determine the coupling $\<\mathcal{O}\mathcal{O}[TT]_{n,\ell}\>$ from the four-point function $\<\mathcal{O}\mathcal{O}\mathcal{O}\mathcal{O}\>$ we need to find a Witten diagram which has a two-graviton cut \cite{Fitzpatrick:2011dm,Yuan:2017vgp,Meltzer:2019nbs}. This first happens at one-loop and one universal contribution is given in figure \ref{fig:universal_graviton_loop}.
\\

The loop diagram on the left appears in every gravitational theory since the coupling $\<\mathcal{O}\mathcal{O} T\>$ is fixed and non-zero by CFT Ward identities. Since the diagram has an internal, $s$-channel, two graviton cut, this implies $[TT]_{n,\ell}$ is exchanged in this channel \cite{Fitzpatrick:2011dm,Fitzpatrick:2015qma,Aharony:2016dwx,Meltzer:2019nbs}. On the other hand, if we cut the internal $\phi$ lines, i.e. perform a $t$-channel cut, we get a product of tree-level Witten diagrams for $\<\mathcal{O}\mathcal{O}\mathcal{O}\mathcal{O}\>$. Therefore, this diagram can be bootstrapped purely from the scalar correlator, which we will review in more detail momentarily.

Once we allow for non-minimal couplings between the scalar and the graviton, we see there are new diagrams with two-graviton cuts. For example, if the $\<TT\mathcal{O}\>$ coupling is non-zero at tree-level in $1/C_T$, we can have a box diagram involving just internal gravitons, as shown in figure \ref{fig:purely_graviton_loop}. There is no way to cut this diagram such that it factorizes into two-scalar, four-point diagrams. Therefore, this loop {\it cannot} be determined purely from $\<\mathcal{O}\mathcal{O}\mathcal{O}\mathcal{O}\>$ at tree-level but instead we are forced to study $\<\mathcal{O}\mathcal{O}TT\>$.\foot{We can be more general: in figure \ref{fig:purely_graviton_loop} we can also replace two gravitons, e.g. the two horizontal lines, with a generic, massive, spin-two particle and get another allowed box diagram with a two-graviton cut.}

\begin{figure}[h]
\begin{center}
\includegraphics[scale=.25]{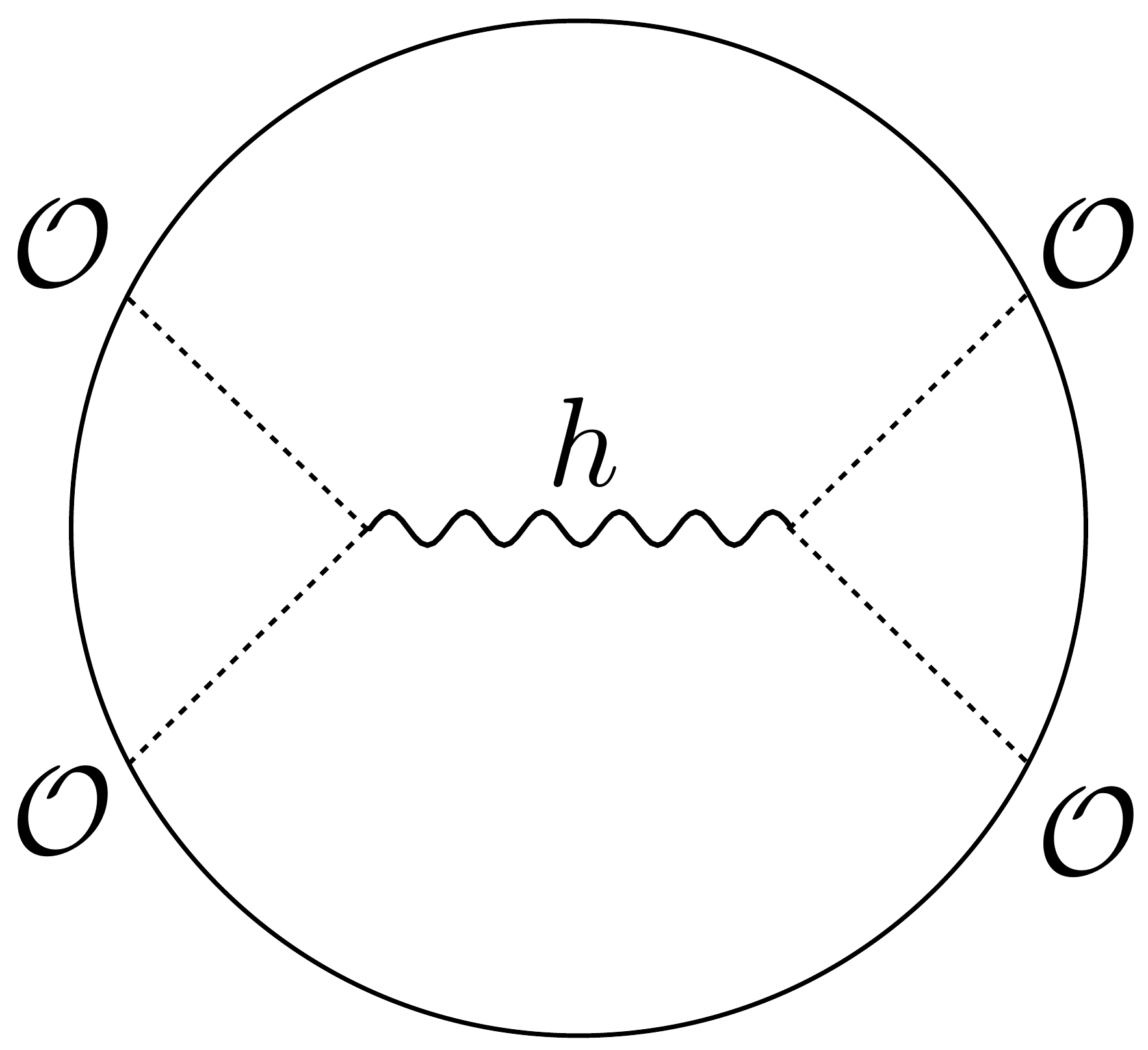}
\end{center}
\caption{}
\label{fig:s-channel_graviton}
\end{figure}

To explain how this works in more detail we review some basic facts about the inversion formula. For the correlator $\<\mathcal{O}\mathcal{O}\mathcal{O}\mathcal{O}\>$ we can write the OPE function $c^{t}(\Delta,J)$ as an integral of the correlator \cite{Caron-Huot:2017vep}:
\begin{align}
c^{t}(\Delta,J)&=\frac{\kappa_{\Delta+J}}{4}\int\limits_{0}^{1}dz d\bar{z}\left |\frac{z-\bar{z}}{z\bar{z}}\right|^{d-2}\frac{1}{(z\bar{z})^{2}}g_{J+d-1,\Delta+1-d}(z,\bar{z})\text{dDisc}_{t}[\mathcal{G}(z,\bar{z})]. \label{eq:inversion}
\\
\kappa_{\beta}&=\frac{\Gamma\left(\frac{\beta}{2}\right)^{4}}{2\pi^{2}\Gamma(\beta-1)\Gamma(\beta)},
\end{align}
Here the reduced correlator $\mathcal{G}$ and $t$-channel double-discontinuity are defined by
\begin{align}
\<\mathcal{O}\mathcal{O}\mathcal{O}\mathcal{O}\>=\frac{1}{(z\bar{z})^{2\Delta_{\mathcal{O}}}}\mathcal{G}(z,\bar{z}) \ , ~~~ 
\text{dDisc}_{t}[\mathcal{G}(z,\bar{z})]=\mathcal{G}(z,\bar{z})-\frac{1}{2}\left(\mathcal{G}^{\circlearrowleft}(z,\bar{z})+\mathcal{G}^{\circlearrowright}(z,\bar{z})\right),
\end{align}
where we are working in the conformal frame (\ref{eq:conformal_frame}) and the arrows indicate how we analytically continue $\bar{z}$ around the branch cut at $\bar{z}=1$. The relation to the physical OPE data is
\begin{align}
\lambda_{\mathcal{O}\mathcal{O}\chi}^{2}=-\res\limits_{\Delta=\Delta_{\chi}} c^{(t)}(\Delta,J_{\chi})(1+(-1)^{J_{\chi}}).
\end{align}
To compute dDisc$_{t}$ of the box diagram in figure \ref{fig:universal_graviton_loop} we need to cut the internal $\phi$ lines, which here give a product of graviton exchange diagrams \cite{Meltzer:2019nbs}. Equivalently, at the level of the OPE we can compute the same double-discontinuity by expanding $\mathcal{G}(z,\bar{z})$ in (\ref{eq:inversion}) in conformal blocks and keeping only the double-trace operators $[\mathcal{O}\mathcal{O}]$. In particular, we should only include their anomalous dimensions due to tree-level graviton exchange. To see this, we should recall that taking dDisc$_{t}$ of $\mathcal{G}(z,\bar{z})$ simply weights each $t$-channel block by a $\sin^2$ factor:
\begin{align}
\text{dDisc}_{t}[\mathcal{G}(z,\bar{z})]=\left(\frac{z\bar{z}}{(1-z)(1-\bar{z})}\right)^{\Delta_{\mathcal{O}}}\sum\limits_{\Delta,J} 2\sin^{2}\left(\frac{\pi}{2}(\Delta-2\Delta_{\mathcal{O}})\right)\lambda_{\Delta,J}^{2} g_{\Delta,J}(1-z,1-\bar{z}). \label{eq:dDiscsum}
\end{align}
The dimensions of the double-trace operators $[\O\O]_{n,\ell}$ take the form $2\Delta_{\O}+2n+\ell + \gamma_{n,\ell}$, where $\gamma_{n,\ell}$ is their anomalous dimensions, so they first start contributing to the dDisc$_{t}$ at one-loop, or $1/C_{T}^2$, and are proportional to their squared anomalous dimensions. In the bulk, the exchange of the double-traces $[\O\O]$ comes from Witten diagrams with a two-particle $\phi$ cut. Here, the relevant anomalous dimensions come from decomposing an $s$-channel graviton exchange diagram $\mathcal{W}^{(s)}_{T}$, see figure \ref{fig:s-channel_graviton}, in terms of $t$-channel blocks. This is of course the diagram which appears when performing a $t$-channel cut of figure \ref{fig:universal_graviton_loop}.
\\

Studying this tree-level Witten diagram gives the crossing equation
\begin{align}
\mathcal{W}^{(s)}_{T}(x_i)=\sum\limits_{n,\ell} \left(2\lambda^{(0)}_{n,\ell}\lambda^{(1)}_{n,\ell}+(\lambda^{(0)}_{n,\ell})^{2}\gamma^{(1)}_{n,\ell}\partial_{\Delta}\right)g_{n,\ell}(x_i),
\end{align}
where we have expanded the OPE coefficients and anomalous dimensions $\gamma_{n,\ell}$ for the double-trace operators $[\O\O]_{n,\ell}$ at large $C_{T}$
\begin{align}
\lambda_{n,\ell}=\lambda^{(0)}_{n,\ell}+\frac{\lambda^{(1)}_{n,\ell}}{C_{T}}+..., \qquad 
\gamma_{n,\ell}=\frac{\gamma^{(1)}_{n,\ell}}{C_{T}}+...
\end{align}
Finally, if we take these anomalous dimensions, from expanding just the $s$-channel graviton exchange diagram into $t$-channel blocks, and plug it into the double-discontinuity (\ref{eq:dDiscsum}), we recover the corresponding discontinuity of the loop diagram in figure \ref{fig:universal_graviton_loop}.\footnote{The exact mapping between the CFT OPE data and the bulk loop expansion can be found in \cite{Meltzer:2019nbs}. There internal scalar lines were considered, but it generalizes straightforwardly to spinning fields up to possible contact-term ambiguities in the spinning bulk propagators.} From the inversion formula, we can use this to reconstruct the full loop diagram, and in particular its $s$-channel two graviton cuts, up to some contact term ambiguities. Of course, at tree-level we also have $t$- and $u$- channel exchange diagrams which can also be used to construct box diagrams. 

\begin{figure}[h]
\begin{center}
\includegraphics[scale=.27]{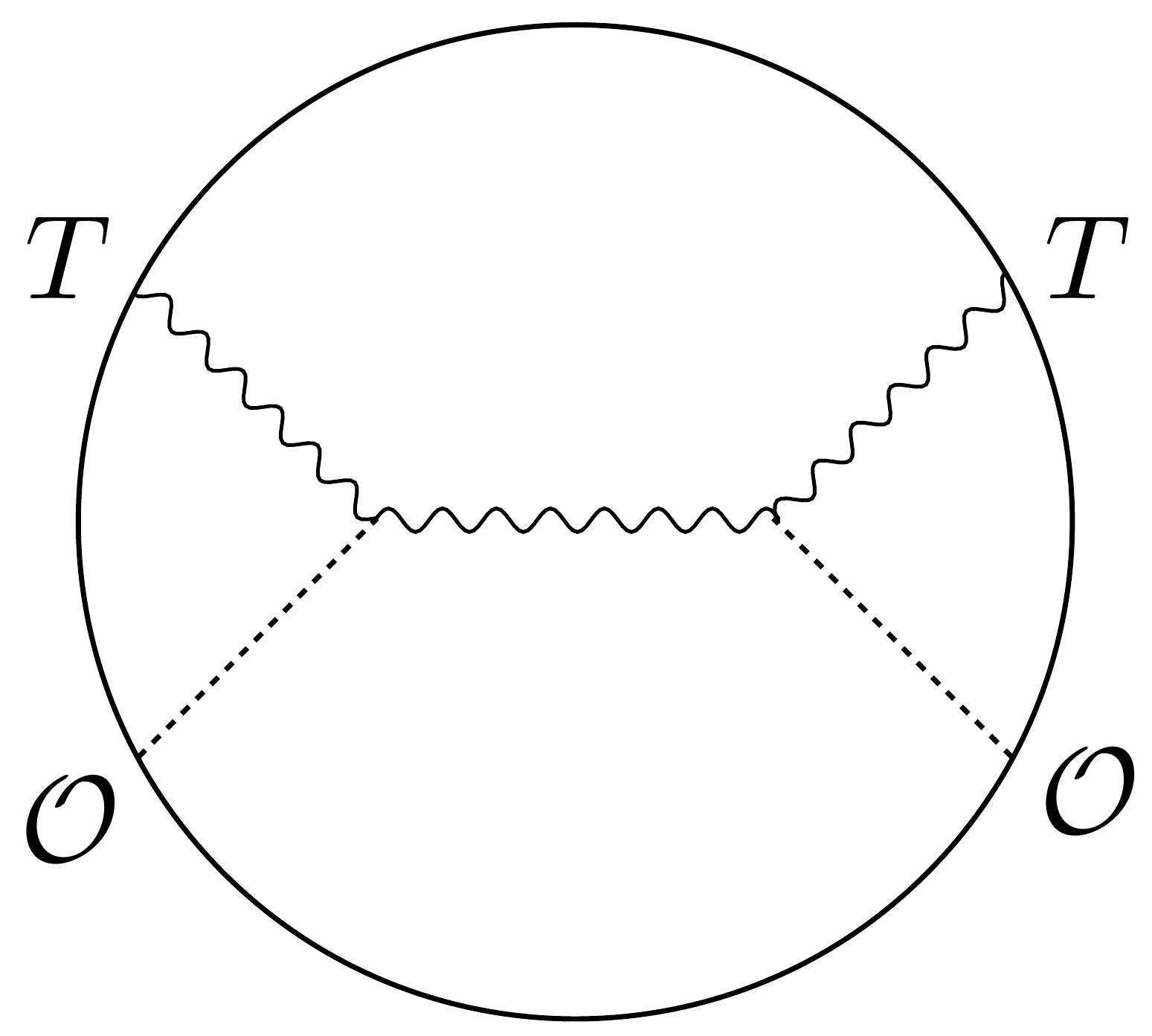}
\end{center}
\caption{}
\label{fig:s-channel_gravitonTTOO}
\end{figure}

The generalization to the new graph in figure \ref{fig:purely_graviton_loop} is now clear. Here if we perform a $t$-channel cut we are left with two graviton exchange diagrams for $\<T\O\O T\>$. Since we are cutting two graviton lines, at the level of the OPE we should be looking for double-traces composed of $T$. Specifically we have the following contribution:
\begin{align}
\text{dDisc}_{t}[\mathcal{G}(z,\bar{z})]\supset\left(\frac{z\bar{z}}{(1-z)(1-\bar{z})}\right)^{\Delta_{\mathcal{O}}}\sum\limits_{n,\ell} 2\sin^{2}\left(\pi(d-\Delta_{\mathcal{O}})\right)\lambda_{\O\O[TT]_{n,\ell}}^{2} g_{[TT]_{n,\ell}}(1-z,1-\bar{z}). \label{eq:dDiscsumTT}
\end{align}
Here we have set the dimension of $[TT]_{n,\ell}$ to its value at $C_T\rar\i$ since the $\sin$ term does not vanish. This is still a one-loop contribution because the individual OPE coefficients, $\lambda_{\O\O[TT]_{n,\ell}}$, scale like $1/C_{T}$. Of course, here we need to be careful about what we mean by ``$\lambda_{\O\O[TT]_{n,\ell}}$" since as discussed in section \ref{s3}, by studying $\<T\O\O T\>$ directly, these OPE coefficients get contributions from multiple possible sources. By cutting figure \ref{fig:purely_graviton_loop}, we should project onto the contribution of the $s$-channel graviton exchange, shown in figure \ref{fig:s-channel_gravitonTTOO}, to $\lambda_{\O\O[TT]_{n,\ell}}$. 

\begin{figure*}[h]
    \centering
    \begin{subfigure}[t]{0.4\textwidth}
        \centering
        \includegraphics[scale=.27]{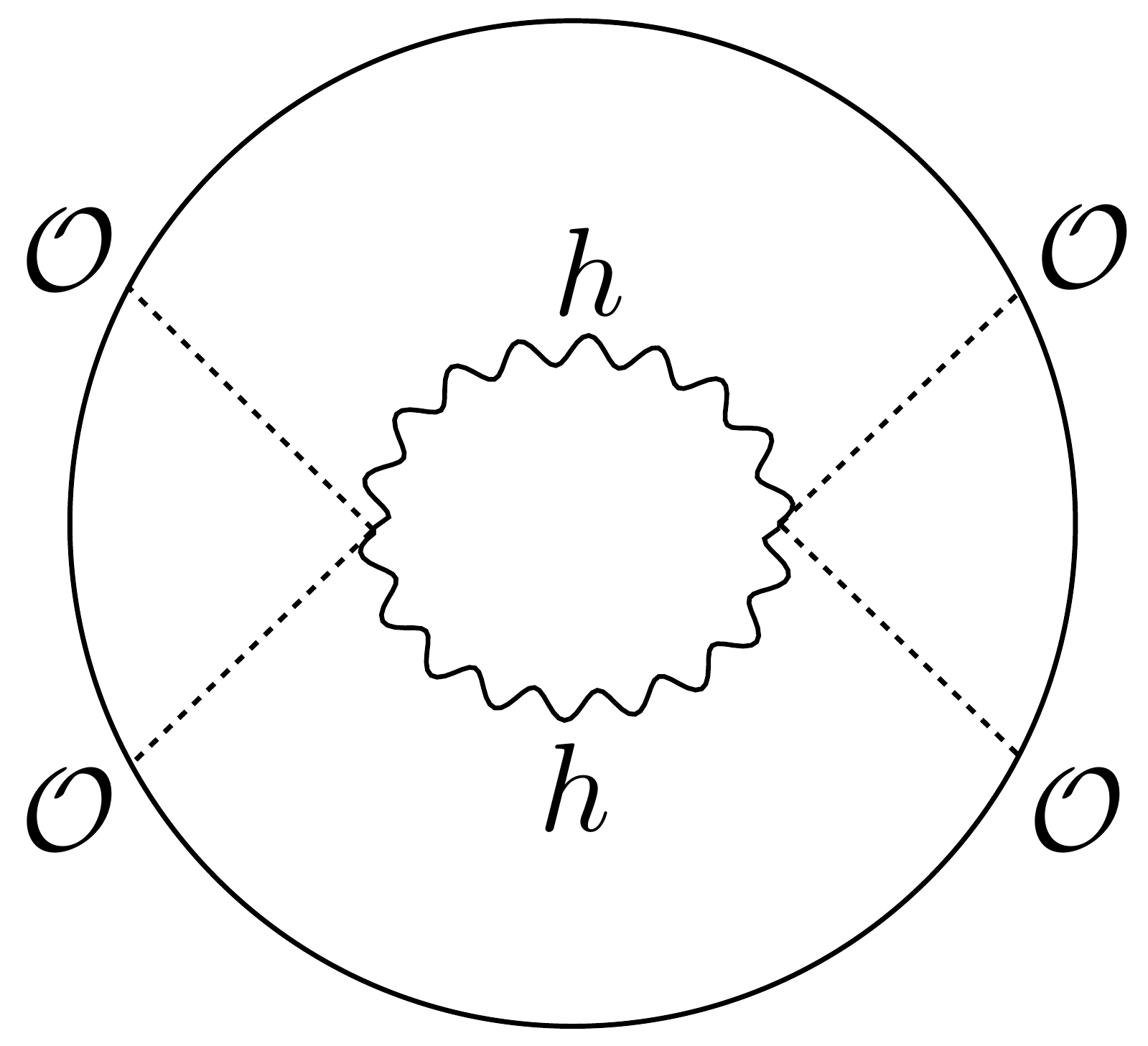}
        \caption{}
       
    \end{subfigure}
    ~ 
    \begin{subfigure}[t]{0.4\textwidth}
        \centering
        \includegraphics[scale=.27]{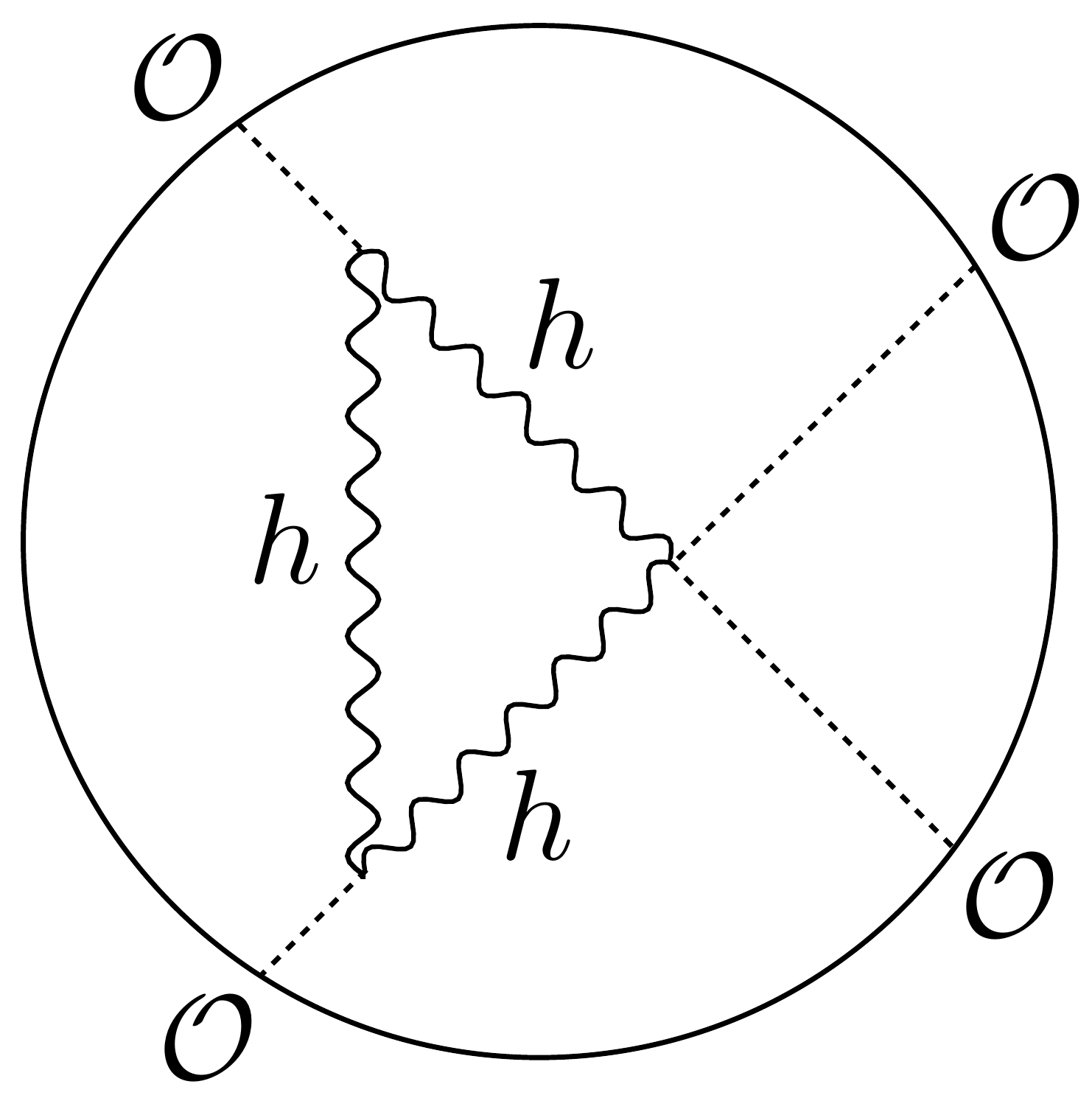}
        \caption{}
        
    \end{subfigure}
    \caption{}
     \label{fig:quartic_loop}
\end{figure*}

This discussion generalizes straightforwardly when we have quartic interactions, which give rise to bubble and triangle diagrams that also have a two-graviton cut, as shown in figure \ref{fig:quartic_loop}.

It is important to emphasize that the existence of a two-graviton cut in these loop diagrams does not necessarily imply that when we perform a conformal block decomposition we find a new contribution to $\<\O\O [TT]_{0,\ell}\>$. One can imagine there are special cancellations for the minimal-twist OPE data. However, the analysis in this work makes it clear that non-minimal cubic and quartic couplings between the scalar and the graviton can affect the tree-level OPE coefficients $\<\mathcal{O}\mathcal{O} [TT]_{0,\ell}\>$ and therefore by unitarity the internal cuts of these loop diagrams. In the language of the bootstrap, we first need to know all tree-level correlators $\<\O\O \O_1\O_2\>$ for arbitrary $\O_{1,2}$ before we can fully determine $\<\O\O\O\O\>$ at one-loop.

\bibliographystyle{utphys}
\bibliography{MTRef}

\end{document}